\documentclass[onecolumn,twoside]{IEEEtran}
\usepackage[T1]{fontenc}
\usepackage[latin9]{inputenc}
\usepackage{amsmath,amssymb,amsfonts}
\usepackage{amsthm}
\usepackage{mathtools}
\usepackage{graphicx} 
\usepackage{cite}
\usepackage{algorithm}
\usepackage{algpseudocode}
\usepackage{multirow}
\usepackage{makecell}
\usepackage{diagbox}
\usepackage{tabularx}
\usepackage{array}
\usepackage[normalem]{ulem}
\usepackage{tikz}
\usepackage{tikz-3dplot}
\usetikzlibrary{arrows.meta, automata, positioning}
\usepackage{comment}
\usepackage{subfigure}
\usepackage{esint}
\usepackage{epstopdf}
\usepackage{float}
\usepackage{lipsum}
\usepackage{balance}
\usepackage{bbm}
\newtheorem{theorem}{Theorem}

\newtheorem{lemma}{Lemma}
\newtheorem{definition}{Definition}
\newtheorem{remark}{Remark}
\newtheorem{corollary}{Corollary}
\newtheorem{example}{Example}

\IEEEoverridecommandlockouts
\definecolor{green}{rgb}{0.13, 0.55, 0.13}
\definecolor{brown}{rgb}{0.6, 0.2, 0.0}
\begin{document}
\title{Current Should Not Sneak: Constrained Codes for Reliable Memristor Crossbar Arrays}

\author{
   \IEEEauthorblockN{Selahattin Kaan K{\i}rge\c{c}, Yunus Alp B{\i}y{\i}ko\u{g}lu, and Ahmed Hareedy, \IEEEmembership{Member, IEEE}}
   \thanks{This work was supported in part by the T\"{U}B\.{I}TAK 2232-B International Fellowship for Early Stage Researchers.
   
    Selahattin Kaan K{\i}rge\c{c} and Ahmed Hareedy are with the Department of Electrical and Electronics Engineering, Middle East Technical University (METU), 06800 Ankara, Turkey (e-mail: kirgec.kaan@metu.edu.tr; ahareedy@metu.edu.tr).

    Yunus Alp B{\i}y{\i}ko\u{g}lu is with the Department of Electrical and Computer Engineering, University of Illinois Urbana-Champaign (UIUC), Urbana, IL 61801 USA (e-mail: yunusab2@illinois.edu).}\vspace{-1.5em}
}
\maketitle

\begin{abstract}

The approach of squeezing more transistors in the same area in order to speed up computing is no longer effective. Currently, researchers and engineers are searching for novel solutions that offer faster computing. One of these solutions is to compute where you store, commonly known as in-memory computing, and it also addresses parallel processing challenges. Resistive random access memories (ReRAMs), which are based on memristor crossbar arrays, enable in-memory computing. Moreover, ReRAMs offer large storage capacity associated with energy efficiency. In this work, we focus on storing digital data in memristor crossbar arrays. A critical challenge here is the sneak-path problem, occurring when there is a rectangle on the array with three low and one high resistances at the corners. The electric current in this case is prone to sneaking through the low-resistance path upon reading, which results in the high resistance data becoming erroneous. In this paper, we propose effective constrained coding solutions to the sneak-path problem after finding the expected number of sneak paths over a two-dimensional array given their circumferences. In particular, we adopt a literature model where $b$ rows on the crossbar array are read simultaneously while the others are grounded, and we design capacity-achieving non-binary constrained codes for the cases of $b=2$ and $b=3$. We focus more on the sneak paths with shorter circumferences as they are more detrimental. Here, GF refers to Galois field. Our GF$(4)$ codes, for $b=2$, and GF$(8)$ codes, for $b=3$, are a class of lexicographically-ordered constrained (LOCO) codes, and we call them resistive-LOCO (RES-LOCO) codes. RES-LOCO codes operate horizontally, and we also suggest a run-length-limited scheme for coding data on the crossbar array vertically to mitigate the sneak-path problem for $b=4$. We experimentally demonstrate the effectiveness of our RES-LOCO codes in remarkably reducing the number of sneak paths, and we offer numerical results for various array setups.

\end{abstract}


\section{Introduction}\label{sec_intro}

In today's world, fast and efficient computing is necessary for a wide range of applications, including data mining, autonomous vehicles, healthcare, and edge devices. For example, the meteoric rise of machine learning algorithms requires speedy distributed computing. The world has witnessed the steady decline of Moore's law, and the fundamental idea of shrinking the device size to gain computing power has become a thing of the past. Instead, parallel computing has been the answer for some time, and a notable example is the usage of multi-core graphical processing units (GPUs) to support multi-core central processing units (CPUs). A major hurdle facing distributed computing is the data transfer back-and-forth between the computing and the storage units, which is called the von Neumann bottleneck \cite{von_neumann}. A promising solution to this bottleneck is in-memory computing, and the relevant systems we focus on here are resistive random access memories (ReRAMs) based on memristor crossbar arrays. In-memory computing enables processing to occur directly where data is stored. Moreover, ReRAMs offer remarkably large storage capacities, and they are energy-efficient.

The memristor, an abbreviation of memory resistor, was first introduced by Chua in 1971 as the missing circuit element \cite{Chua1971Memristor}. Between voltage, current, flux linkage, and charge, Chua introduced the memristor to relate flux and charge, which was the missing link. Since then, the physics of memristors was extensively studied. Physical memristor realization was discussed in \cite{Strukov2008Missing}. A version of Ohm's law that is state dependent was introduced in \cite{Chua2019Resistance} to define a memristor. Charge-based vs resistance-based memristor devices were compared in \cite{ref1}. Here, we focus on using memristor crossbar arrays in digital storage as a first step towards in-memory computing. While the memristor offers two stable states for binary storage, it suffers from a serious problem called the sneak-path problem. The sneak-path problem, in brief, occurs when a rectangle on the two-dimensional crossbar array has three low resistances, each representing logic $1$, and one high resistance, representing logic $0$, at its four corners. Naturally, the electric current takes the parallel path with three low resistances in this case, resulting in the likelihood of the logic $0$ mistakenly read as logic $1$ upon reading \cite{Zidan2013Memristor}.

From the circuits perspective, there is also a rich literature discussing memristors and their crossbar arrays. A detailed overview of the materials used, switching techniques, device modeling, and performance analysis was presented in \cite{Zahoor2020Resistive}. The effect of wordline and bitline scaling on the performance and energy efficiency of the crossbar array was discussed in \cite{Liang2013Effect}. Circuit-based solutions to the sneak-path problem were also introduced in \cite{Zidan2013Memristor} and in \cite{Lee2025Recent}. Memristors and ReRAMs have several different applications. The usage of memristors for in-memory computing was discussed in \cite{ref3}. How memristors can be adopted for faster deep learning, efficient neural networks, as well as neuromorphic computing was illustrated in \cite{ref2}. A survey of the past, present, and future of memristors can be found in \cite{Chen2020ReRAM}.

From the coding and information theoretic perspective, there are many recent results for memristor crossbar arrays. There are results focusing on constrained coding for error prevention, and others focusing on error-correction coding. Cassuto et al. modeled the associated channel as a z-channel where an error can only occur from a $0$ to a $1$ \cite{Cassuto2014Channel}. The same group also introduced a novel idea to mitigate sneak-path effects by reading a number $b$ of rows (wordlines) simultaneously at any given time while grounding all other crossbar array rows \cite{Cassuto2016Information}, and they offered constrained coding solutions. A bridge from memristor physics-circuits to memristor information-coding theoretic techniques was built in \cite{Dupraz2023Turning}. Nguyen et al. presented a straightforward sneak-path-mitigating approach based on run-length-limited (RLL) constrained codes \cite{nguyen2023locally}. Maximum a posteriori (MAP) detection and constrained coding were introduced in \cite{BenHur2019Detection}. On the side of error correction, there are techniques based on low-density parity-check (LDPC) codes \cite{Pang2025Across}, polar codes \cite{ref6}, as well as irregular repeat-accumulate (IRA) codes \cite{Song2024Performance} for ReRAMs and their crossbar arrays.

There are other data processing approaches customized for memristor crossbar arrays. A communications-inspired sneak-path estimation approach was proposed in \cite{Naous2014Memristor}. Chen et al. introduced a technique for adaptive reading and detection to mitigate sneak-path effects \cite{Chen2019Pilot}. Rate coding with memristors for spiking neural networks was presented in \cite{Pallathuvalappil2025Rate}. On the error-correction decoding side, there are results on bit-flipping algorithms \cite{Kim2022Sneak}, majority-logic algorithms \cite{Kong2025Sneak}, as well as joint detection and decoding based on belief propagation \cite{Sun2022Belief} to enhance the reliability of ReRAMs.

Constrained codes prevent error-prone data patterns from being written (transmitted) in order to enhance the reliability of data storage (transmission) systems. Shannon introduced these codes in 1948 under the label coding for discrete noiseless systems as he represented the constrained system via a finite-state transition diagram (FSTD) and evaluated the capacity \cite{shan_const}. Constrained codes can be designed using finite-state machines or lexicographic indexing, and we here focus on the latter. Early developments on the design of constrained codes based on lexicographic indexing, also called enumerative codes, include the run-length-limited (RLL) code design by Tang and Bahl in \cite{tang_bahl}. More recent works on enumerative constrained codes include \cite{datta_2}, \cite{Hareedy2019Asymmetric} and \cite{ahh_qaloco} for Flash memory systems, as well as \cite{wang_etal} for DNA data storage. In 2019, Hareedy and Calderbank introduced lexicographically-ordered constrained codes, in short LOCO codes \cite{Hareedy2019LOCO}. Based on Cover's result in \cite{cover_lex}, a general method to design LOCO codes for any finite set of forbidden patterns was then presented in \cite{Hareedy2022LOCO}. LOCO codes achieve capacity with low complexity since their encoding-decoding algorithms are based on a simple mathematical rule that can be executed by a reconfigurable adder \cite{Hareedy2022LOCO}. Since then, this general method was used to design advanced non-binary LOCO codes for modern data storage systems \cite{Ozbayrak2024TDLOCO}, \cite{Irimagzi2024Protecting}, \cite{reins_EC_DLOCO}.

In general, we adopt the model in \cite{Cassuto2016Information}, where $b$ rows are read simultaneously on the crossbar array while grounding all other rows, in this work. However, we also show that our schemes notably enhance ReRAM reliability under other models. Our contribution in this paper is four-fold:
\begin{enumerate}
\item We compute the expected number of sneak paths over any two-dimensional (2D) crossbar array for a specific circumference and for a given circumference range as well as for a given probability of low resistance. We show that such expectations offer accurate estimates of the actual sneak-path counts.

\item For the case of $b=2$, we design effective and efficient GF($4$) LOCO codes to remove all sneak paths of all circumferences within each group of two rows. Here, GF refers to Galois field. We call our proposed codes resistive-LOCO codes, in short RES-LOCO codes.

\item For the case of $b=3$, we design effective and efficient GF($8$) LOCO codes to remove all sneak paths of the shortest circumference, which is $4$, within each group of three rows. Such sneak paths are known to be the most detrimental \cite{Chen2019Pilot}.

\item For the case of $b=4$, we propose a binary RLL scheme for coding data on the crossbar array vertically to mitigate the sneak-path problem.
\end{enumerate}
RES-LOCO codes operate horizontally on the rows. For RES-LOCO codes, we convert a two-dimensional binary problem into a one-dimensional non-binary problem. We offer the detailed analysis of these codes and show that they naturally achieve capacity. All our proposed coding schemes offer feasible code rates for data storage. RES-LOCO codes are simple and can be easily reconfigured as the ReRAM device ages because of their encoding-decoding rule. Over different crossbar array setups and grounding mechanisms, we experimentally show that the proposed coding schemes significantly reduce the number of sneak paths, notably increasing the reliability of the ReRAM device.

The rest of the paper is organized as follows. In Section~\ref{sec_count}, we introduce the necessary preliminaries and estimate sneak-path counts. In Section~\ref{sec_gf4}, we design our GF($4$) RES-LOCO codes for memristor crossbar arrays. In Section~\ref{sec_gf8}, we design our GF($8$) RES-LOCO codes for memristor crossbar arrays. In Section~\ref{sec_rates}, we discuss finite-length challenges and state the code rates. In Section~\ref{sec_rll}, we present our binary RLL constrained coding idea. In Section~\ref{sec_sims}, we introduce our experimental results. In Section~\ref{sec_conc}, we conclude the paper and state future work.

\section{Preliminaries and Count Estimation}\label{sec_count}

Memristor-based ReRAMs, designed using crossbar arrays, are becoming popular since they can store a large amount of data in a small area and consume less energy than traditional memory technologies. They also enable in-memory computing, where data can be processed in the same hardware where it is stored \cite{ref1,ref2,ref3}. However, these crossbar arrays suffer from the \textit{sneak-path problem}, which causes read errors. This problem is discussed below in detail.

To mitigate the sneak-path problem, researchers have proposed various coding and circuit strategies, although these strategies can still be improved to further increase the overall storage capacity \cite{ref6}. Our approach employs LOCO codes, which encode and decode data in a simple way that avoids these harmful configurations. By applying LOCO codes to ReRAM crossbar arrays, we can eliminate the most detrimental sneak paths and approach capacity. In fact, LOCO codes achieve capacity under a given constraint. Next, we discuss the sneak-path problem.

\begin{figure}
\centering
\tdplotsetmaincoords{60}{125} 

\begin{tikzpicture}[
    tdplot_main_coords,
    scale=1.2,
    wl/.style={line width=2.5pt, color=cyan!70!blue}, 
    bl/.style={line width=2.5pt, color=blue!70!black}, 
    pillar/.style={line width=1.2pt, color=gray!90}, 
    mem_low/.style={fill=green!40!black, draw=white, line width=1pt, circle, inner sep=2.5pt},
    mem_high/.style={fill=red!80!black, draw=white, line width=1pt, circle, inner sep=2.5pt},
    expected/.style={-{Latex[length=3mm, width=2mm]}, line width=1.8pt, color=green!55!black, solid, rounded corners=3pt},
    sneak/.style={-{Latex[length=3mm, width=2mm]}, line width=1.8pt, color=red!85!black, dashed, rounded corners=3pt}
]

\def\D{1.8}     
\def\Zgap{1.8}  
\def\Zmem{0.5}  
\def\off{0.25}  

\foreach \x in {1, 2, 3} {
    \draw[bl] (\x*\D, 0.2*\D, 0) -- (\x*\D, 4.2*\D, 0);
    \node[right, color=blue!70!black, font=\small\bfseries] at (\x*\D, 0.24*\D, 0.2) {BL$_{\x}$}; 
}

\foreach \x in {1,2,3} {
    \foreach \y in {3,2,1} {
        
        \draw[pillar] (\x*\D, \y*\D, 0) -- (\x*\D, \y*\D, \Zgap);
        
        \fill[black] (\x*\D, \y*\D, 0) circle (1.2pt);       
        \fill[black] (\x*\D, \y*\D, \Zgap) circle (1.2pt);   
        
        \def\memcolor{mem_low} 
        \ifnum\y=3 
            \ifnum\x=2 \def\memcolor{mem_high} \fi 
            \ifnum\x=3 \def\memcolor{mem_high} \fi 
        \fi
        \ifnum\y=2 
            \ifnum\x=1 \def\memcolor{mem_high} \fi 
            \ifnum\x=2 \def\memcolor{mem_high} \fi 
        \fi
        \ifnum\y=1 
            \ifnum\x=1 \def\memcolor{mem_high} \fi 
        \fi
        
        \node[\memcolor] at (\x*\D, \y*\D, \Zmem) {};
    }
}

\foreach \y in {1, 2, 3} {
    \draw[wl] (0.2*\D, \y*\D, \Zgap) -- (4.2*\D, \y*\D, \Zgap);
    \node[left, color=cyan!70!blue, font=\small\bfseries] at (0.2*\D, \y*\D, 1.1*\Zgap) {WL$_{\y}$};
}


\draw[expected] (0.5*\D, 2*\D+\off, \Zgap) 
             -- (2*\D+\off, 2*\D+\off, \Zgap) 
             -- (2*\D+\off, 2*\D+\off, 0) 
             -- (2*\D+\off, 0.5*\D, 0);

\node[above right, color=green!55!black, font=\small\bfseries] 
    at (1.5*\D, 2.3*\D, \Zgap) {expected};

\draw[sneak] (0.5*\D, 2*\D-\off, \Zgap) 
          -- (3*\D-\off, 2*\D-\off, \Zgap) 
          -- (3*\D-\off, 2*\D-\off, 0)
          -- (3*\D-\off, 1*\D-\off, 0)
          -- (3*\D-\off, 1*\D-\off, \Zgap) 
          -- (2*\D-\off, 1*\D-\off, \Zgap)
          -- (2*\D-\off, 1*\D-\off, 0) 
          -- (2*\D-\off, 0.5*\D, 0);   

\node[right, color=red!85!black, font=\small\bfseries] 
    at (1*\D-\off, -1*\D, 0) {sneak path};

\node[mem_low, label=right:{\footnotesize\bfseries \begin{tabular}{c}Low R\\(1)\end{tabular}}] 
    at (0.3*\D, 5*\D, 0) {};
    
\node[mem_high, label=right:{\footnotesize\bfseries \begin{tabular}{c}High R\\(0)\end{tabular}}] 
    at (1.5*\D, 5*\D, 0) {};

\end{tikzpicture}
\caption{Memristor crossbar array showing expected current path (green solid) and sneak path (red dashed).}
\label{fig1}
\end{figure}
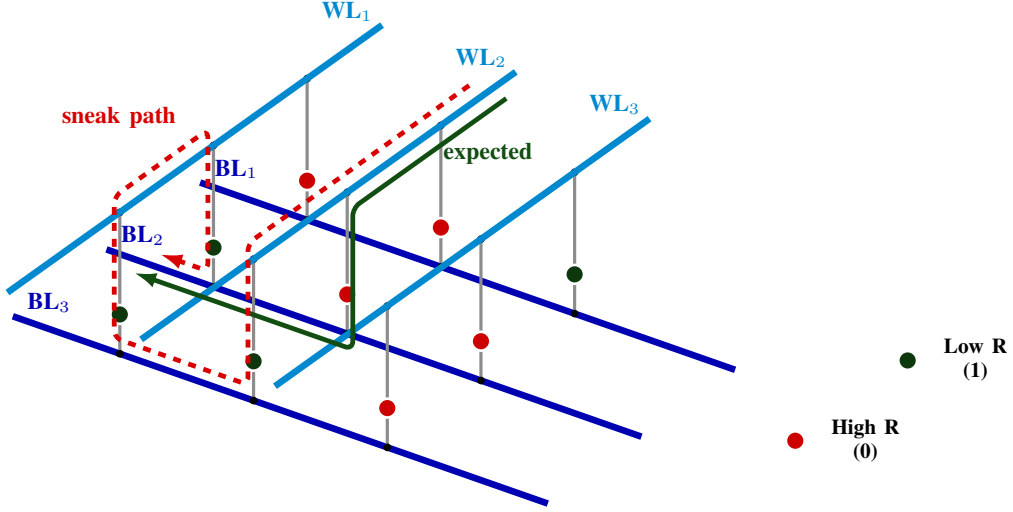

In ReRAM-based crossbar architectures, one of the main challenges is the emergence of unintended current pathways, called \textit{sneak paths}, which may arise during both read and write operations~\cite{Cassuto2016Information}. In this work, we primarily focus on the read operation problem. Before we formally define and express the sneak path mathematically, consider Fig.~\ref{fig1}. Memristor cells exist at each wordline-bitline intersection. In this figure, high resistance (low resistance) cells are shown as red (green) small circles. Suppose we want to read the cell at the intersection of Wordline $2$ and Bitline $2$. Ideally, the current should follow the expected line, which is green solid, in the figure. However, due to low resistances at $(2,3)$, $(1,3)$, and $(1,2)$, where the notation is wordline index followed by bitline index, the current may instead follow the sneak-path line, which is red dashed. Consequently, a high-resistance cell representing a $0$ can be mistakenly read as a low-resistance cell representing a $1$ due to the sneak-path problem. Therefore, we can say that a sneak path exists at $(2,2)$.

\begin{definition}[Sneak-Path Problem]\label{def:sneak_path}
In a crossbar array, let $a(i, j)$ denote the stored binary value of the memristor at position $(i, j)$. High (Low) resistance, which implies low (high) conductance, corresponds to $a(i, j)=0$ ($a(i, j)=1$). A high resistance cell at position $(i, j)$, i.e., $a(i, j) = 0$, is subject to a sneak path if there exist indices $s \neq i$ and $c \neq j$ such that there are three low resistance cells as follows:
\begin{equation} \label{def_sp}
a(i, c) = a(s, j) = a(s, c) = 1,
\end{equation}
which create an alternative current route $(i, c) \to (s, c) \to (s, j)$ that bypasses the target cell during the read operation. The circumference of this sneak path is $L = 2(|s - i| + |c - j|)$.
In an array with no selectors and no diodes (1R array), this path is always active. In a 1D1R array, each cell has a diode that blocks unwanted current. In this case, the sneak path is active only when the diode at the corner cell $(s, c)$ fails \cite{Chen2019Pilot}. In a 1S1R array, each cell has a selector that exhibits high resistance if the cell is on a possible sneak path. In this case, the sneak path is active only when the selectors at all three rectangular sneak-path corners, other than $(i, j)$, fail \cite{Chen2019Pilot}.
\end{definition}

\begin{remark}
In the absence of a diode or a selector, sneak-path configurations are more general and cannot be restricted to rectangular shapes. However, in 1D1R (one diode-one memristor) and 1S1R (one selector-one memristor) architectures, the formation of non-rectangular paths would require simultaneous failure of multiple diodes, more in the case of 1S1R, which has a statistically negligible probability. Moreover, the dominance of longer sneak paths is significantly mitigated by the cumulative effect of line resistance, making them notably less detrimental. Consequently, the following assumptions are maintained throughout this work:
\begin{enumerate}
    \item Sneak paths are assumed to be strictly rectangular. In 1D1R or 1S1R architectures, this implies the failure of a single diode or three selectors, respectively. In crossbar arrays without diodes/selectors, only rectangular forms are considered, while others are ignored.
    \item Following Definition~\ref{def:sneak_path}, only paths up to a determined maximum circumference are considered. Longer paths are ignored as increasing line resistance reduces their impact on system performance.
\end{enumerate}
\end{remark}

To address the sneak-path problem, we propose LOCO coding schemes. To evaluate the effectiveness of our coding schemes, we derive the expected number of sneak paths first, which is a result we will use later. In this section, we derive general formulae for the expected number of sneak paths with arbitrary circumferences, then construct various LOCO coding schemes that improve performance in the following sections.

\begin{lemma}[Expected Number of Sneak Paths With Specific Circumference]\label{lemma:exp_sp_l}
Consider a $(\gamma_1 \times \gamma_2)$ crossbar array where memristor states are independent and identically distributed, with $p_1$ denoting the probability that a memristor is in the low-resistance state ($1$ in binary). For sufficiently large dimensions $\gamma_1$ and $\gamma_2$, the expected number of sneak paths with circumference $L = 2\ell + 4$, where $\ell \in \{0, 1, 2, \ldots,\min(\gamma_1,\gamma_2)-2\}$, is given by:
\begin{equation}\label{lemma_1}
\begin{aligned}
A(\ell) = \frac{2}{3}p_1^3(1-p_1)(\ell+1)\Big[6\gamma_1\gamma_2-3(\gamma_1+\gamma_2)(\ell+2)+(\ell+2)(\ell+3)\Big],
\end{aligned}
\end{equation}
where $\ell$ is the effective circumference parameter, i.e., actual circumferences $L \in \{4, 6, 8, \ldots\}$ are mapped to $\ell \in \{0, 1, 2, \ldots\}$, and $A(\ell)$ is the expected number of sneak paths with circumference $L$.
\begin{proof}
To find $A(\ell)$, based on Definition~\ref{def:sneak_path}, we probabilistically analyze all valid placements of rectangular shapes with circumference $L$ in the $(\gamma_1 \times \gamma_2)$ crossbar array. While the sneak path technically refers to the path of the current, we also use the same terminology to refer to the rectangular shape causing it to occur.

First, we consider the probability of forming a sneak path for a fixed set of coordinates, i.e., fixed four corners. Since we require three memristors to be in the low-resistance state (LRS) and one to be in the high-resistance state (HRS), there are $\binom{4}{1} = 4$ such configurations. The probability for any specific rectangle location is therefore $4p_1^3(1 - p_1)$.

Next, we determine the number of valid spatial positions of these rectangular shapes. Let $r$ and $c$ represent the lengths of the row and column spans of the path, respectively. The circumference constraint $L=2\ell+4$ implies that $2(r+c) = 2\ell+4$, which simplifies to $r+c = \ell+2$. For a fixed pair of lengths $(r, c)$, the number of ways to embed this rectangle into a $(\gamma_1 \times \gamma_2)$ array is $(\gamma_1 - r)(\gamma_2 - c)$.

Summing over all valid integer combinations of $r$ and $c$ gives:
\[
A(\ell) = 4p_1^3(1 - p_1) \sum_{\mathclap{\{(r,c) \, | \, r+c=\ell+2, \, (r,c) \in \{1,2,\ldots,\ell+1\}^2\}}}(\gamma_1 - r)(\gamma_2 - c).
\]
To evaluate this sum, we substitute $c = (\ell + 2) - r$. Since $r \geq 1$ and $c \geq 1$, the summation index $r$ ranges from $1$ to $\ell+1$. Therefore, the expression becomes:
\[
A(\ell) = 4p_1^3(1 - p_1) \sum_{r=1}^{\ell+1} (\gamma_1 - r)\Big(\gamma_2 - (\ell + 2) + r\Big).
\]
Let $K = \ell + 2$. The term inside the summation can be expanded as a polynomial in $r$ as follows:
\[
(\gamma_1-r)(\gamma_2 - K + r) = (\gamma_1\gamma_2 - \gamma_1K) + r(\gamma_1 - \gamma_2 + K) - r^2.
\]
We now apply the known summation identities for the first $n$ integers and their squares, where $n = \ell+1$:
\begin{equation*}
\sum_{r=1}^{n} r = \frac{n(n+1)}{2}, \quad \sum_{r=1}^{n} r^2 = \frac{n(n+1)(2n+1)}{6}.
\end{equation*}
Substituting these identities into the expanded sum and simplifying the resulting algebraic expression yields the final expression of $A(\ell)$ in \eqref{lemma_1} and completes the proof.
\end{proof}
\end{lemma}

\begin{lemma}[Cumulative Expected Number of Sneak Paths]\label{lemma:exp_sp_l_max}
Consider an $(\gamma_1 \times \gamma_2)$ crossbar array where memristor states are independent and identically distributed, with $p_1$ denoting the probability that a memristor is in the low-resistance state ($1$ in binary). For sufficiently large dimensions $\gamma_1$ and $\gamma_2$, the cumulative expected number of sneak paths with circumferences ranging from $L = 4$ to $L_{\max} = 2\ell_{\max} + 4$, where $\ell_{\max} \in \{0, 1, 2, \ldots, \min(\gamma_1,\gamma_2)-2\}$, is given by:
\begin{align}
B(\ell_{\max}) &= \frac{p_1^3 (1 - p_1)}{6} \big[
\gamma_1\gamma_2 \big( 24 + 36\ell_{\max} + 12\ell_{\max}^2 \big)
- (\gamma_1 + \gamma_2) \big( 24 + 44\ell_{\max} \nonumber \\
&\qquad + 24\ell_{\max}^2 + 4\ell_{\max}^3 \big)
+ \big( 24 + 50\ell_{\max} + 35\ell_{\max}^2 + 10\ell_{\max}^3 + \ell_{\max}^4 \big) \big], \label{lemma_2}
\end{align}
where $\ell_{\max}$ is the maximum effective circumference parameter and $B(\ell_{\max})$ is the cumulative expected number of sneak paths with circumferences from $L = 4$ to $L_{\max}$.
\begin{proof}
When $\min(\gamma_1,\gamma_2) \geq \ell_{\max}+2$, the cumulative count $B(\ell_{\max})$ is obtained by summing the result from Lemma~\ref{lemma:exp_sp_l} over all valid integer parameters $\ell$ as follows:
\[
B(\ell_{\max}) = \sum_{\ell=0}^{\ell_{\max}} A(\ell).
\]
From Lemma~\ref{lemma:exp_sp_l}, we know that $A(\ell)$ is a polynomial in $\ell$ of degree $3$. In particular, $A(\ell)$ contains terms involving $\ell^0$, $\ell^1$, $\ell^2$, and $\ell^3$. To evaluate the cumulative sum, we rely on the linearity of the summation operator and known power sum identities up to the third degree. In addition to the identities of $\sum \ell$ and $\sum \ell^2$ used in the proof of Lemma~\ref{lemma:exp_sp_l}, we utilize the identity of the sum of integer cubes:
\vspace{-0.1em}\[
\sum_{\ell=0}^{n} \ell^3 = \left( \frac{n(n+1)}{2} \right)^2,
\]
where $n = \ell_{\max}$.
By substituting the explicit form of $A(\ell)$ from \eqref{lemma_1} into the summation of $B(\ell_{\max})$ and applying these identities to the terms with powers of $\ell$, we obtain the algebraic expression in \eqref{lemma_2}.
\end{proof}
\end{lemma}

The lemmas we have introduced provide a general model for sneak-path enumeration in arrays sufficiently large to include all possible sneak-path configurations (e.g., a circumference-sixteen sneak path occurring as $2(1+7)$, $2(2+6)$, etc.). For finite dimensions, direct expressions can be found. For example, consider a $2 \times 100$ memristor array. Here, we can express the sneak-path expectation contribution as $99K$ for circumference-four paths, $98K$ for circumference-six paths, $97K$ for circumference-eight paths, and so on, where $K = 4p_1^3(1 - p_1)$ is the probabilistic multiplier. For any memristor model, the shortest sneak paths are considered the most dominant. This is because they take up less physical space, making them more common in a limited area. Moreover, in larger sneak paths, the current encounters significantly more line resistance, which weakens the sneak-path effect~\cite{Liang2013Effect} as the effective resistance of the possible sneak path increases. This observation~\cite{Chen2019Pilot} and our lemmas hold true for general crossbar arrays.

Throughout this paper, we use the general methodology for constructing LOCO codes as described in~\cite{Hareedy2022LOCO}. The goal is to find the encoding-decoding function $g(\mathbf{c})$, which gives the lexicographic index of any codeword $\mathbf{c}\triangleq c_{m-1}c_{m-2}\dots c_{0}$ in the codes defined in the following sections. The steps of this general methodology can be summarized as follows:
\begin{enumerate}
  \item \textit{Partition the code into groups based on the forbidden patterns.}
  \item \textit{Formulate the codebook size using the recursive nature of the group hierarchy.}
  \item \textit{Characterize the special-case codeword patterns arising from the forbidden pattern constraints.}
  \item \textit{Derive the lexicographic index contribution of each non-zero symbol for both special and typical cases.}
  \item \textit{Synthesize a unified index equation from all case contributions, establishing the LOCO encoding-decoding rule.}
  \item \textit{Design the encoding and decoding algorithms that implement this rule.}
\end{enumerate}

\section{RES-LOCO Coding Scheme Over \text{GF}$(4)$}\label{sec_gf4}

In this section, we first introduce the GF$(4)$ RES-LOCO coding scheme that prevents sneak paths of circumference $4$. Then, we develop the GF$(4)$ coding scheme for general circumference range from $4$ to $L_{\max}$ with increments of $2$. These codes are defined over GF$(4) = \{0,1,\alpha,\alpha^2\}$, and they forbid the sneak-path (SP) patterns where the row span is always $1$. We adopt the following mapping-demapping between GF$(4)$ and $2$-tuple array columns in order to specify the forbidden patterns in a way that eliminates the relevant SP patterns. We then design and analyze the constrained codes.

The GF$(4)$ RES-LOCO coding scheme adopts the following $\text{GF}(4) \longleftrightarrow [\text{GF}(2)]^2$ mapping-demapping:
\begin{align}\label{eqn_gf4map}
0 &\longleftrightarrow [ 0 ~ 0]^\mathrm{T}, \hspace{+3.6em} 1 \longleftrightarrow [0 ~ 1 ]^\mathrm{T}, \nonumber \\
\alpha &\longleftrightarrow [1 ~ 0 ]^\mathrm{T}, \hspace{+3.0em} \alpha^2 \longleftrightarrow [1 ~ 1 ]^\mathrm{T}.
\end{align}
We adopt the general methodology in \cite{Hareedy2022LOCO} for the construction of RES-LOCO codes. The process begins with an examination of the forbidden patterns, from which the cardinality equations are derived, followed by the steps leading to deriving the encoding-decoding rule. Our forbidden patterns for preventing sneak paths of circumference $4$ are in
\begin{equation}\label{GF(4)_l_0_forbid}
    \mathcal{T}^{4,0} = \{ 1\alpha^2, \alpha\alpha^2, \alpha^2 1, \alpha^2 \alpha \}.
\end{equation}
The set corresponds to the $[1 ~ 1 ]^\mathrm{T}$ column associated from either side with a column that has exactly one $0$, which creates a sneak path. The generalization of this set for any $\ell$, denoted by $\mathcal{T}^{4,\ell}$, is in \eqref{GF(4)_l_general_forbid}.

\begin{remark}
A key feature of our design is that it is independent of any specific memristor model, and the channel analysis in \cite{Chen2019Pilot} supports our approach. Consequently, the proposed solution is applicable to any memristor architecture such as 1R, 1D1R, and 1S1R memristor crossbar arrays, although its performance may vary depending on physical parameters such as wire resistance and diode/selector endurance.
\end{remark}

The formal definition of a GF$(4)$ RES-LOCO code is as follows:
\begin{definition}[GF$(4)$ RES-LOCO Code]\label{def:GF(4) RES-LOCO}
A RES-LOCO code, $\mathcal{RES}^{4,\ell}_{m}$, is defined by the following properties:
\begin{enumerate}
    \item Codewords in $\mathcal{RES}^{4,\ell}_{m}$ are defined over $\textup{GF}(4)$, the code alphabet, and are of length $m$ symbols.
    \item Codewords in $\mathcal{RES}^{4,\ell}_{m}$ are lexicographically ordered.
    \item Codewords in $\mathcal{RES}^{4,\ell}_{m}$ do not contain any patterns from the set $\mathcal{T}^{4,\ell}$.
     \item Any codeword satisfying the above properties is included in $\mathcal{RES}^{4,\ell}_{m}$ .
\end{enumerate}
\end{definition}
Lexicographic ordering means codewords are ordered following the notion $0 < 1 < \alpha < \alpha^2$ and symbol significance reduces from left to right within the codeword. A codeword $\mathbf{c}$ is defined as the sequence $c_{m-1} c_{m-2} \dots c_ 1 c_0$. Now, we develop the $\mathcal{RES}^{4,0}_{m}$ code to illustrate the main ideas on a simpler code, and then we develop the RES-LOCO code for general $\ell$. When we say ``starting'' for a codeword, we always mean from the left.

\textit{First, we specify the group structure.} Groups of $\mathcal{RES}^{4,0}_{m}$ codes that prevents sneak paths of circumference $L = 4$ are:
\begin{itemize}
    \item Group~$1$ contains all the codewords starting with 
          $0x$, where $x$ can be any of the four GF$(4)$ symbols.
    \item Group~$2$ contains all the codewords starting with 
          $1\delta_1$, where $\delta_1 \in \{0,1,\alpha\}$.
    \item Group~$3$ contains all the codewords starting with 
          $\alpha\delta_1$, where $\delta_1 \in \{0,1,\alpha\}$.
    \item Group~$4$ contains all the codewords starting with 
          $\alpha^2\delta_2$, where $\delta_2 \in \{0,\alpha^2\}$.
\end{itemize}

\textit{Second, we enumerate the codewords.}
Let $N_{4,0}(m)$ denote the number of $\mathcal{RES}^{4,0}_{m}$ codewords, which have length $m$. Additionally, let $N_{4,0,i}(m)$ denote the number of $\mathcal{RES}^{4,0}_{m}$ codewords belonging to Group~$i$, where $i\in\{1, 2, 3, 4\}$. From this, it follows that $N_{4,0}(m) = \sum_{i = 1}^{4}N_{4,0,i}(m)$.
\begin{theorem}\label{thm:N(m)_for_GF4_l0}
The cardinality $N_{4,0}(m)$ of a RES-LOCO code $\mathcal{RES}^{4,0}_{m}$, for $m\geq3$, is given by:
\begin{equation}
    N_{4,0}(m)=4N_{4,0}(m-1)-2N_{4,0}(m-2)-2N_{4,0}(m-3),\label{eq:N(m)_for_GF4_l0}
\end{equation}
where the defined cardinalities are $N_{4,0}(2)\triangleq12\text{, }N_{4,0}(1)\triangleq4 \text{, and }N_{4,0}(0)\triangleq1$.
\begin{proof}
    From the group structure, it can be seen that 
    \begin{equation}\label{thm1,eq1}
        N_{4,0,1}(m) = N_{4,0}(m-1),
    \end{equation}
    since any codeword in $\mathcal{RES}^{4,0}_{m-1}$ can be concatenated to the right of the symbol $0$ for Group~$1$. Similarly, we have the relation
    \begin{equation}\label{thm1,eq2}
        N_{4,0,2}(m) = N_{4,0,3}(m) = N_{4,0}(m-1)-N_{4,0,4}(m-1),
    \end{equation}
    since any codeword in $\mathcal{RES}^{4,0}_{m-1}$ starting with a symbol in $\{0, 1, \alpha\}$ can be concatenated to the right of the symbol $1$ for Group~$2$ or to the symbol $\alpha$ for Group~$3$. Additionally,
    \begin{equation}\label{thm1,eq3}
    \begin{aligned}
        N_{4,0,4}(m) = N_{4,0,1}(m-1) + N_{4,0,4}(m-1)&\\= N_{4,0}(m-2) +N_{4,0,4}(m-1).
    \end{aligned}
    \end{equation}
    By using $N_{4,0}(m) = \sum_{i = 1}^{4}N_{4,0,i}(m)$ along with \eqref{thm1,eq1}, \eqref{thm1,eq2}, and \eqref{thm1,eq3}, we reach
    \begin{equation}\label{thm1,eq4}
        N_{4,0}(m) = 3N_{4,0}(m-1)+N_{4,0}(m-2)-N_{4,0,4}(m-1).
    \end{equation}
    Since we know that $N_{4,0,4}(m-1)-N_{4,0,4}(m-2) = N_{4,0}(m-3)$ from \eqref{thm1,eq3}, if we find $N_{4,0}(m)-N_{4,0}(m-1)$ via \eqref{thm1,eq4}, we reach \eqref{eq:N(m)_for_GF4_l0}. Moreover, $N_{4,0}(1)$ and $N_{4,0}(2)$ can be found directly using the set of forbidden patterns. As for $N_{4,0}(0)$, we can use \eqref{thm1,eq1} because $N_{4,0,1}(1) = N_{4,0}(0) = 1$.
\end{proof}
\end{theorem}

\textit{Third and Fourth, we find the special cases and symbol contribution.}
After deriving the necessary cardinality relations, we are able to advance to the next step, which is to define the typical and special cases to be used when calculating the contributions of symbols. We define the contribution of a given symbol $c_i$ to the overall codeword index $g(\mathbf{c})$ as the number of smaller-length codewords that we can generate by replacing that symbol with $c'_i < c_i$ and the symbols to its right with valid constrained alternatives. Valid here means we can concatenate them from the right to $c_{m-1} c_{m-2} \dots c_{i+2} c_{i+1}$, the symbols of $\mathbf{c}$, without violating the constraint. To calculate the contribution of symbol $c_i$, we must look at the preceding symbols $c_{m-1} \dots c_{i+2} c_{i+1}$. This is necessary because replacing $c_i c_{i-1} \dots c_0$ with shorter-length codewords as illustrated above might create a forbidden pattern once concatenated to $c_{m-1} \dots c_{i+2} c_{i+1}$. If this concatenation creates a forbidden pattern, which we characterize as a ``special'' case, some of the shorter-length codewords need to be omitted to correctly calculate the contribution of the symbol at hand. Otherwise, all shorter-length codewords can be concatenated and counted in the symbol contribution, which we characterize as a ``typical'' case. We then find the symbol contribution for each case.

\begin{theorem}\label{res_gf4_0_rule}
The encoding-decoding rule of a RES-LOCO code $\mathcal{RES}^{4,0}_{m}$, for $m\geq3$, is given by:
\begin{equation}
g(\mathbf{c}) = \sum_{i=0}^{m-1} \Big [ k_{i,1} [N_{4,0}(i+1) - N_{4,0}(i-1)] + k_{i,2} N_{4,0}(i) \Big ],
\end{equation}
where $k_{i,1}$ and $k_{i,2}$ are defined according to the special/typical case of symbol $c_i$ in $\mathbf{c}$ as shown in \eqref{symb_cont_1}.

\begin{proof}
Using the forbidden patterns that we previously specified in \eqref{GF(4)_l_0_forbid}, we identify the typical and special cases as well as determine their corresponding symbol contributions $g_i(c_i)$ as follows:
\begin{itemize}
\item Special case $S_1$, $c_{i+1}c_i=\alpha^2\alpha^2$: $g_i(c_i) = N_{4,0,1}(i+1) = N_{4,0}(i)$ since only codewords starting with $c'_i = 0$ in $\mathcal{RES}^{4,0}_{i+1}$ can contribute because of the forbidden patterns.
\item Typical case $T_1$, $c_{i+1}c_i=\delta 1$, $\delta \in \{0, 1, \alpha\}$: $g_i(c_i) = N_{4,0,1}(i+1) = N_{4,0}(i)$.
\item Typical case $T_2$, $c_{i+1}c_i=\delta\alpha$: $g_i(c_i) = N_{4,0,1}(i+1) + N_{4,0,2}(i+1) = N_{4,0}(i+1)-N_{4,0}(i)-N_{4,0}(i-1)$.
\item Typical case $T_3$, $c_{i+1}c_i=0\alpha^2$: $g_i(c_i) = N_{4,0,1}(i+1) + 2N_{4,0,2}(i+1) = 2N_{4,0}(i+1)-3N_{4,0}(i)-2N_{4,0}(i-1)$.
\end{itemize}
In the above equations of $g_i(c_i)$, we use the group cardinalities in Theorem~\ref{thm:N(m)_for_GF4_l0} and its proof, including that $N_{4,0}(i+1)=4N_{4,0}(i)-2N_{4,0}(i-1)-2N_{4,0}(i-2)$. If $c_i$ corresponds to the leftmost symbol of the codeword, then we assume that $c_{i+1} = c_m \triangleq 0$ for the purposes of identifying typical and special cases for $c_{m-1}$ (always typical).

\textit{Fifth, we formulate the encoding-decoding rule.}
After outlining the symbol contributions for each special and typical case to the overall codeword index, we are now able to find the merged indexing rule, which combines all the aforementioned cases. We define indicator functions $y_{i,S_1}$, $y_{i,T_1}$, $y_{i,T_2}$, and $y_{i,T_3}$ that attain the value $1$ whenever the current case is $S_1$, $T_1$, $T_2$, or $T_3$, respectively, and have the value $0$ otherwise. With these indicator functions, we write the merged indexing rule as follows:

\begin{align}\label{symb_cont_1}
    g_i(c_i) &= \underbrace{(y_{i,T_2}+2y_{i,T_3})}_{k_{i,1}} \, [N_{4,0}(i+1) - N_{4,0}(i-1)] + \underbrace{(y_{i,T_1}-y_{i,T_2}-3y_{i,T_3}+y_{i,S_1})}_{k_{i,2}} N_{4,0}(i) \nonumber \\
             &= k_{i,1} [N_{4,0}(i+1) - N_{4,0}(i-1)] + k_{i,2} N_{4,0}(i),
\end{align}
\begin{equation}\label{gen_rule_1}
g(\mathbf{c}) = \sum_{i=0}^{m-1}g_i(c_i).
\end{equation}
Combining \eqref{symb_cont_1} and \eqref{gen_rule_1} completes the proof and gives the encoding-decoding rule.
\end{proof}
\end{theorem}

Next, we introduce examples to clarify the idea of the encoding-decoding rule. From Theorem~\ref{thm:N(m)_for_GF4_l0}, we find that $N_{4,0}(3) = 38$.

\begin{example}[Decoding]\label{ex_decode}
We find the index of the codeword $\alpha^2\alpha^2\alpha^2$ in $\mathcal{RES}_3^{4,0}$. For $i = 0$ and $i = 1$, we identify the special case $S_1$. For $i = 2$, we identify the typical case $T_3$. Using the encoding-decoding rule:
\begin{align*}
g(\alpha^2\alpha^2\alpha^2) &= g_0(c_0) + g_1(c_1) + g_2(c_2) \\
&= N_{4,0}(0) + N_{4,0}(1) + [2N_{4,0}(3) - 3N_{4,0}(2) - 2N_{4,0}(1)] \\
&= 1 + 4 + [2(38) - 3(12) - 2(4)] = 37.
\end{align*}
This result is consistent with $N_{4,0}(3) = 38$, as $\alpha^2\alpha^2\alpha^2$ is the last codeword of length $3$.
\end{example}

\begin{example}[Encoding]\label{ex_encode}
We find the codeword with index $36$ in $\mathcal{RES}_3^{4,0}$. For $i = 2$, the possible contributions are:
\begin{itemize}
    \item $T_1$: $g_2(1) = N_{4,0}(2) = 12$.
    \item $T_2$: $g_2(\alpha) = N_{4,0}(3) - N_{4,0}(2) - N_{4,0}(1) = 22$.
    \item $T_3$: $g_2(\alpha^2) = 2N_{4,0}(3) - 3N_{4,0}(2) - 2N_{4,0}(1) = 32$.
    \item $S_1$: $g_2(\alpha^2) = N_{4,0}(2) = 12$.
\end{itemize}
We select the largest contribution that does not exceed $36$, which is $g_2(\alpha^2) = 32$ associated with $T_3$. Thus, $c_2 = \alpha^2$ and the remaining index is $36 - 32 = 4$.

For $i = 1$, given that $c_2 = \alpha^2$, we identify the special case $S_1$ with $g_1(\alpha^2) = N_{4,0}(1) = 4$. Thus, $c_1 = \alpha^2$ and the remaining index becomes $4 - 4 = 0$.

For $i = 0$, since the remaining index is $0$, we have $c_0 = 0$.

Therefore, the codeword with $g(\mathbf{c})=36$ is $\alpha^2\alpha^20$. This result is consistent with our decoding example, as the index of $\alpha^2\alpha^2\alpha^2$ is $37$, confirming that $\alpha^2\alpha^20$ immediately precedes it.
\end{example}

Now, we extend the GF$(4)$ RES-LOCO coding scheme to prevent sneak paths of all circumferences $L \in \{4, 6, 8, \ldots, L_{\max}\}$, where $L_{\max} = 2\ell + 4$ and $\ell \in \{0, 1, 2, \ldots\}$. We adopt the general methodology in \cite{Hareedy2022LOCO} for designing the RES-LOCO code. The process begins with an examination of the forbidden patterns, using which the cardinality equations are derived, followed by the derivation of the encoding and decoding rule.

Our forbidden patterns for preventing sneak paths of circumferences up to $L_{\max}$ are in
\begin{align}\label{GF(4)_l_general_forbid}
\mathcal{T}^{4,\ell} = \{&1\alpha^2, 10\alpha^2, 1\mathbf{0}^2\alpha^2, \ldots, 1\mathbf{0}^\ell\alpha^2,  \alpha\alpha^2, \alpha0\alpha^2, \alpha\mathbf{0}^2\alpha^2, \ldots, \alpha\mathbf{0}^\ell\alpha^2, \nonumber \\
&\alpha^21, \alpha^201, \alpha^2\mathbf{0}^21, \ldots, \alpha^2\mathbf{0}^\ell1, \alpha^2\alpha, \alpha^20\alpha, \alpha^2\mathbf{0}^2\alpha, \ldots, \alpha^2\mathbf{0}^\ell\alpha\}.
\end{align}

Here, we adopt the notation $\mathbf{y}^r$ to denote a run of $r$ consecutive identical symbols $y$, i.e., $\mathbf{y}^r = \overbrace{y\,y\,\cdots\,y}^{r}$, with $\mathbf{y}^0$ denoting the empty string.

\textit{First, we specify the group structure.} Groups of $\mathcal{RES}^{4,\ell}_{m}$ code are:
\begin{itemize}
    \item Group~$1$ contains all the codewords starting with 
          $0x$, where $x$ can be any of the four GF$(4)$ symbols.
    \item Group~$2$ contains all the codewords starting with $1$:
    \begin{itemize}
        \item Subgroup~$2_{k,1}$: $1\mathbf{0}^k\alpha\delta_1$, where $k = 0, 1, \ldots, \ell$, $\delta_1 \in \{0, 1, \alpha\}$.
        \item Subgroup~$2_{k,2}$: $1\mathbf{0}^k1\delta_1$, where $k = 0, 1, \ldots, \ell$.
        \item Subgroup~$2_{\ell+1}$: $1\mathbf{0}^{\ell+1}$.
    \end{itemize}
    \item Group~$3$ contains all the codewords starting with $\alpha$:
    \begin{itemize}
        \item Subgroup~$3_{k,1}$: $\alpha\mathbf{0}^k\alpha\delta_1$, where $k = 0, 1, \ldots, \ell$, $\delta_1 \in \{0, 1, \alpha\}$.
        \item Subgroup~$3_{k,2}$: $\alpha\mathbf{0}^k1\delta_1$, where $k = 0, 1, \ldots, \ell$.
        \item Subgroup~$3_{\ell+1}$: $\alpha\mathbf{0}^{\ell+1}$.
    \end{itemize}
    \item Group~$4$ contains all the codewords starting with $\alpha^2$:
    \begin{itemize}
        \item Subgroup~$4_k$: $\alpha^2\mathbf{0}^k\alpha^2\delta_2$, where $k = 0, 1, \ldots, \ell$, $\delta_2 \in \{0, \alpha^2\}$.
        \item Subgroup~$4_{\ell+1}$: $\alpha^2\mathbf{0}^{\ell+1}$.
    \end{itemize}
\end{itemize}

\textit{Second, we enumerate the codewords.}
Let $N_{4,\ell}(m)$ denote the number of $\mathcal{RES}^{4,\ell}_{m}$ codewords, which have length $m$. Additionally, let $N_{4,\ell,i}(m)$ denote the number of $\mathcal{RES}^{4,\ell}_{m}$ codewords belonging to Group~$i$, where $i\in\{1, 2, 3, 4\}$. From this, it follows that $N_{4,\ell}(m) = \sum_{i = 1}^{4}N_{4,\ell,i}(m)$.

\begin{theorem}\label{thm:N(m)_for_GF4_l_general}
The cardinality $N_{4,\ell}(m)$ of a RES-LOCO code $\mathcal{RES}^{4,\ell}_{m}$, for $m\geq 2\ell+3$, is given by:
\begin{equation}\label{eq:N(m)_for_GF4_l_general}
    N_{4,\ell}(m)=4N_{4,\ell}(m-1)-2\sum_{k=2}^{2\ell+3}N_{4,\ell}(m-k),
\end{equation}
where the defined cardinalities can be computed based on the group structure for the value of $\ell$ of interest.
\begin{proof}
    From the group structure, we observe that 
    \begin{equation}\label{thm2,eq1}
        N_{4,\ell,1}(m) = N_{4,\ell}(m-1),
    \end{equation}
     since any codeword in $\mathcal{RES}^{4,\ell}_{m-1}$ can be concatenated to the right of the symbol $0$ for Group~$1$. 
    
For the subgroups of Group~$2$, and also Group~$3$ by symmetry, we have
\begin{align}\label{thm2,eq2}
    N_{4,\ell,2_{k,1}}(m) + N_{4,\ell,2_{k,2}}(m) + N_{4,\ell,4_k}(m) &= N_{4,\ell}(m-k-1) - N_{4,\ell,1}(m-k-1) \nonumber \\ &= N_{4,\ell}(m-k-1) - N_{4,\ell}(m-k-2),
\end{align}
for $k = 0, 1, \ldots, \ell$. The reason is that the codewords of $\mathcal{RES}^{4,\ell}_{m}$ in these subgroups are all the codewords in $\mathcal{RES}^{4,\ell}_{m-k-1}$ that do not start with a $0$.

Using \eqref{thm2,eq2} and summing over all subgroups of Groups~$2$ and $4$ collectively yields
\begin{align}
    N_{4,\ell,2}(m) + N_{4,\ell,4}(m) &= \sum_{k=0}^{\ell}[N_{4,\ell}(m-k-1) - N_{4,\ell}(m-k-2)] + N_{4,\ell,2_{\ell+1}}(m) + N_{4,\ell,4_{\ell+1}}(m) \nonumber \\ &= \sum_{k=0}^{\ell}[N_{4,\ell}(m-k-1) - N_{4,\ell}(m-k-2)] + 2N_{4,\ell}(m-\ell-2) \nonumber \\
    &= N_{4,\ell}(m-1) + N_{4,\ell}(m-\ell-2). \label{thm2,eq3}
\end{align}
The second equality follows from that codewords in Subgroup~$2_{\ell+1}$ or in Subgroup~$4_{\ell+1}$ of $\mathcal{RES}^{4,\ell}_{m}$ are all codewords in $\mathcal{RES}^{4,\ell}_{m-\ell-1}$ that start with a $0$.
    
    The individual group cardinalities satisfy
    \begin{align}
        N_{4,\ell,2}(m) &= N_{4,\ell}(m-1) - \sum_{k=1}^{\ell+1}N_{4,\ell,4}(m-k), \label{thm2,eq4}\\
        N_{4,\ell,4}(m) &= N_{4,\ell}(m-1) - 2\sum_{k=1}^{\ell+1}N_{4,\ell,2}(m-k). \label{thm2,eq5}
    \end{align}
As for \eqref{thm2,eq4}, the justification is that codewords in Group~$2$ at length $m$ are all codewords at length $m-1$ except those starting with $\alpha^2$, $0\alpha^2$, $\mathbf{0}^2\alpha^2$, \dots, or $\mathbf{0}^\ell \alpha^2$, which can also be obtained from Group~$4$ at lengths $m-1$, $m-2$, \dots, or $m-\ell-1$. As for \eqref{thm2,eq5}, the justification is that codewords in Group~$4$ at length $m$ are all codewords at length $m-1$ except those starting with $\beta$, $0\beta$, $\mathbf{0}^2\beta$, \dots, or $\mathbf{0}^\ell \beta$, where $\beta \in \{1,\alpha\}$, which can also be obtained from Groups~$2$ and $3$ at lengths $m-1$, $m-2$, \dots, or $m-\ell-1$.
    
    Since $N_{4,\ell}(m) = N_{4,\ell,1}(m) + 2N_{4,\ell,2}(m) + N_{4,\ell,4}(m)$, using \eqref{thm2,eq1}, \eqref{thm2,eq4}, and \eqref{thm2,eq5}, we obtain
    \begin{equation}\label{thm2,eq6}
        N_{4,\ell}(m) = 4N_{4,\ell}(m-1) - 2\sum_{k=1}^{\ell+1}[N_{4,\ell,2}(m-k) + N_{4,\ell,4}(m-k)].
    \end{equation}
    Substituting \eqref{thm2,eq3} for length $m-k$ in \eqref{thm2,eq6} then gives
\begin{align}\label{thm2,eq7}
    N_{4,\ell}(m) &= 4N_{4,\ell}(m-1)
         - 2\sum_{k=1}^{\ell+1}[N_{4,\ell}(m-k-1) + N_{4,\ell}(m-k-\ell-2)] \nonumber \\
         &= 4N_{4,\ell}(m-1) - 2\sum_{k=2}^{2\ell+3}N_{4,\ell}(m-k).
\end{align}
    
    This completes the proof. For further verification, we check the case of $\ell = 0$. Here, \eqref{eq:N(m)_for_GF4_l_general} reduces to $N_{4,0}(m) = 4N_{4,0}(m-1) - 2N_{4,0}(m-2) - 2N_{4,0}(m-3)$, which agrees with \eqref{eq:N(m)_for_GF4_l0}.
\end{proof}
\end{theorem}

For convenience, and since they are also needed in the derivations of the encoding-decoding rule, we use Equations \eqref{thm2,eq1}--\eqref{thm2,eq7} to determine the following recursive relations for the cardinalities of Groups~$2$, $3$, and $4$ in $\mathcal{RES}^{4,\ell}_{m}$:
\begin{align}
N_{4,\ell,2}(m) &= N_{4,\ell,3}(m) = N_{4,\ell}(m) - 2N_{4,\ell}(m-1) - N_{4,\ell}(m-\ell-2). \label{thm3_groups1} \\
N_{4,\ell,4}(m) &= -N_{4,\ell}(m) + 3N_{4,\ell}(m-1) + 2N_{4,\ell}(m-\ell-2). \label{thm3_groups2}
\end{align}

\textit{Third and Fourth, we find the special cases and symbol contribution.}
After deriving the necessary cardinality relations, we are able to advance to the next step, which is to define the typical and special cases to be used when calculating the contributions of symbols. We define the contribution of a given symbol $c_i$ to the overall codeword index $g(\mathbf{c})$ as the number of smaller-length codewords that we can generate by replacing that symbol with $c'_i < c_i$ and the symbols to its right with valid constrained alternatives. Valid here means we can concatenate them from the right to $c_{m-1} c_{m-2} \dots c_{i+2} c_{i+1}$, the symbols of $\mathbf{c}$, without violating the constraint. To calculate the contribution of symbol $c_i$, we must look at the preceding symbols $c_{m-1} \dots c_{i+2} c_{i+1}$. This is necessary because replacing $c_i c_{i-1} \dots c_0$ with shorter-length codewords as illustrated above might create a forbidden pattern once concatenated to $c_{m-1} \dots c_{i+2} c_{i+1}$. If this concatenation creates a forbidden pattern, which we characterize as a ``special'' case, some of the shorter-length codewords need to be omitted to correctly calculate the contribution of the symbol at hand. Otherwise, all shorter-length codewords can be concatenated and counted in the symbol contribution, which we characterize as a ``typical'' case. We then find the symbol contribution for each case.

\begin{theorem}\label{res_gf4_l_rule}
The encoding-decoding rule of a RES-LOCO code $\mathcal{RES}^{4,\ell}_{m}$, for $m\geq3$, is given by:
\begin{align}\label{eqn_gf4_rule}
g(\mathbf{c}) = \sum_{i=0}^{m-1} &\Bigg[ k_{i,1} N_{4,\ell}(i+1) + k_{i,2} N_{4,\ell}(i) + k_{i,3} N_{4,\ell}(i-\ell-1) \nonumber \\
              &+ \sum_{r=1}^{\ell}\Bigg\{ k_{i,4,r} \big[N_{4,\ell}(i+1) - N_{4,\ell}(i-\ell-1)\big] \nonumber \\
              &+ \sum_{j=i+r-\ell-1}^{i-1} \, \Big[ k_{i,5,r} N_{4,\ell}(j+1)
              + k_{i,6,r} N_{4,\ell}(j)
              + k_{i,7,r} N_{4,\ell}(j-\ell-1)\Big] \Bigg\} \Bigg],
\end{align}
where $k_{i,1}$, $k_{i,2}$, $k_{i,3}$, $k_{i,4,r}$, $k_{i,5,r}$, $k_{i,6,r}$, and $k_{i,7,r}$ are defined according to the special/typical case of symbol $c_i$ in $\mathbf{c}$ as shown in \eqref{symb_cont_gf4_l}.
\begin{proof}
Using the forbidden patterns that we previously specified, we identify the typical and special cases as well as determine their corresponding symbol contributions $g_i(c_i)$ as follows:

\textit{Typical cases:}
\begin{itemize}
\item Typical case $T_1$, $c_{i+1}c_i=\zeta 1$, $i=m-1$, or $c_{i+p}\dots c_i=\mathbf{0}^p 1$, where $c_i = 1$ in both cases, $p \geq \ell$, and $\zeta$ is a symbol that refers to out of codeword bounds:
\begin{equation}
g_i(c_i) = N_{4,\ell,1}(i+1) = N_{4,\ell}(i) \label{thm3_contrib0}
\end{equation}
since only the codewords starting with $c'_i = 0$ in $\mathcal{RES}^{4,\ell}_{i+1}$ can be concatenated from the right to the symbols preceding $c_i$ without violating the constraint, and their count is $N_{4,\ell}(i)$.

\item Typical case $T_2$, $c_{i+1}c_i=\zeta \alpha$, $i=m-1$, or $c_{i+p}\dots c_i=\mathbf{0}^p \alpha$, where $c_i = \alpha$ in both cases, $p \geq \ell$, and $\zeta$ is a symbol that refers to out of codeword bounds:
\begin{equation}
g_i(c_i) = N_{4,\ell}(i+1) - N_{4,\ell}(i) - N_{4,\ell}(i-\ell-1) \label{thm3_contrib1}
\end{equation}
since replacing $c_i = \alpha$ with $c'_i = 1$ contributes the codewords starting with $1$ and replacing $c_i = \alpha$ with $c'_i = 0$ contributes the codewords starting with $0$ that can be concatenated without creating a forbidden pattern. Observe that the number of these codewords is obtained via \eqref{thm2,eq1} and \eqref{thm3_groups1}.

\item Typical case $T_3$, $c_{i+p}\dots c_i=\mathbf{0}^p\alpha^2$, where $c_i = \alpha^2$ and $p > \ell$:
\begin{equation}
g_i(c_i) = 2N_{4,\ell}(i+1) - 3N_{4,\ell}(i) - 2N_{4,\ell}(i-\ell-1) \label{thm3_contrib2}
\end{equation}
since replacing $c_i = \alpha^2$ with each of the lexicographically-smaller symbols $c'_i \in \{0, 1, \alpha\}$ contributes the corresponding valid codewords to concatenate, none of which is restricted because the preceding $\mathbf{0}^p$ run, with $p > \ell$, prevents any forbidden pattern. Observe that the number of these codewords is obtained again via \eqref{thm2,eq1} and \eqref{thm3_groups1}.
\end{itemize}

\textit{Special cases:}
\begin{itemize}
\item Special case $S_1$, $c_{i+r}\dots c_i=1\mathbf{0}^{r-1}1$, where $r \in \{1, 2, \ldots, \ell\}$:
\begin{equation}
g_i(c_i) = N_{4,\ell}(i)
- \sum_{j=i+r-\ell-1}^{i-1} \, \big[-N_{4,\ell}(j+1) + 3N_{4,\ell}(j) + 2N_{4,\ell}(j-\ell-1)\big]
\end{equation}
since this is the typical contribution of $c_i =1$ in \eqref{thm3_contrib0} with a necessary subtraction. The short $\mathbf{0}$ run of length $r-1 < \ell$ following the leading $1$ restricts some of the shorter-length codewords, specifically the ones starting with $\alpha^2$ at various shorter lengths, and the summation adds these omitted codewords. Observe that the number of such omitted codewords is given by \eqref{thm3_groups2}.

\item Special case $S_2$, $c_{i+r}\dots c_i=1\mathbf{0}^{r-1}\alpha$, where $r \in \{1, 2, \ldots, \ell\}$:
\begin{equation}
g_i(c_i) = N_{4,\ell}(i+1) - N_{4,\ell}(i) - N_{4,\ell}(i-\ell-1)
- \sum_{j=i+r-\ell-1}^{i-1} \, \big[-N_{4,\ell}(j+1) + 3N_{4,\ell}(j) + 2N_{4,\ell}(j-\ell-1)\big]
\end{equation}
since this is the typical contribution of $c_i =\alpha$ in \eqref{thm3_contrib1} with a necessary subtraction. The short $\mathbf{0}$ run of length $r-1 < \ell$ following the leading $1$ restricts some of the shorter-length codewords, specifically the ones starting with $\alpha^2$ at various shorter lengths, and the summation adds these omitted codewords.

\item Special case $S_3$, $c_{i+r}\dots c_i=\alpha \mathbf{0}^{r-1}1$, where $r \in \{1, 2, \ldots, \ell\}$:
\begin{equation}
g_i(c_i) = N_{4,\ell}(i)
- \sum_{j=i+r-\ell-1}^{i-1} \, \big[-N_{4,\ell}(j+1) + 3N_{4,\ell}(j) + 2N_{4,\ell}(j-\ell-1)\big]
\end{equation}
since this is the typical contribution of $c_i =1$ in \eqref{thm3_contrib0} with a necessary subtraction. The short $\mathbf{0}$ run following the leading $\alpha$ restricts the shorter-length codewords that start with $\alpha^2$, and the summation adds these omitted codewords.

\item Special case $S_4$, $c_{i+r}\dots c_i=\alpha \mathbf{0}^{r-1}\alpha$, where $r \in \{1, 2, \ldots, \ell\}$:
\begin{equation}
g_i(c_i) = N_{4,\ell}(i+1) - N_{4,\ell}(i) - N_{4,\ell}(i-\ell-1)
- \sum_{j=i+r-\ell-1}^{i-1} \, \big[-N_{4,\ell}(j+1) + 3N_{4,\ell}(j) + 2N_{4,\ell}(j-\ell-1)\big]
\end{equation}
since this is the typical contribution of $c_i =\alpha$ in \eqref{thm3_contrib1} with a necessary subtraction. The short $\mathbf{0}$ run following the leading $\alpha$ restricts the shorter-length codewords that start with $\alpha^2$, and the summation adds these omitted codewords.

\item Special case $S_5$, $c_{i+r}\dots c_i=\alpha^2 \mathbf{0}^{r-1}\alpha^2$, where $r \in \{1, 2, \ldots, \ell\}$:
\begin{equation}
g_i(c_i) = N_{4,\ell}(i)
- 2\sum_{j=i+r-\ell-1}^{i-1} \, \big[N_{4,\ell}(j+1) - 2N_{4,\ell}(j) - N_{4,\ell}(j-\ell-1)\big]
\end{equation}
since this is the typical contribution of $c_i =\alpha^2$ in \eqref{thm3_contrib2} with a necessary subtraction. The short $\mathbf{0}$ run of length $r-1 < \ell$ following the leading $\alpha^2$ restricts some of the shorter-length codewords, specifically the ones starting with $1$ or $\alpha$ at various shorter lengths, and the summation adds these omitted codewords. Observe that the number of such omitted codewords is given by \eqref{thm3_groups1} multiplied by $2$.

\item Special case $S_6$, $c_{i+\ell+1}\dots c_i=\alpha^2 \mathbf{0}^\ell \alpha^2$:
\begin{equation}
g_i(c_i) = N_{4,\ell}(i)
\end{equation}
since the $\mathbf{0}$ run of length exactly $\ell$ is just long enough to satisfy the constraint. Thus, only the codewords starting with $c'_i = 0$ contribute, and their number is $N_{4,\ell}(i)$.
\end{itemize}

Observe that if an index extends beyond the leftmost symbol of the codeword, then we assume that $c_k \triangleq 0$ for all $k \geq m$ for the purposes of identifying typical and special cases.

\textit{Fifth, we formulate the encoding-decoding rule.}
After outlining the symbol contributions for each special and typical case to the overall codeword index, we are now able to find the merged indexing rule, which combines all the aforementioned cases. We define indicator functions $y_{i,T_1}$, $y_{i,T_2}$, $y_{i,T_3}$, $y_{i,S_1,r}$, $y_{i,S_2,r}$, $y_{i,S_3,r}$, $y_{i,S_4,r}$, $y_{i,S_5,r}$, and $y_{i,S_6}$ that attain the value $1$ whenever the current case is the corresponding case, and have the value $0$ otherwise. To simplify the analysis, we define merging flags that combine indicator functions for cases with identical symbol contributions as follows:
\begin{align}
y_{i,1} &= y_{i,T_1} + y_{i,S_6}, \\
y_{i,2} &= y_{i,T_2}, \\
y_{i,3} &= y_{i,T_3}, \\
y_{i,4,r} &= y_{i,S_1,r} + y_{i,S_3,r}, \\
y_{i,5,r} &= y_{i,S_2,r} + y_{i,S_4,r}, \\
y_{i,6,r} &= y_{i,S_5,r}.
\end{align}
With these merging flags, we write the merged indexing rule as follows:
\begin{align}\label{symb_cont_gf4_l}
    g_i(c_i) &= (y_{i,2} + 2y_{i,3}) N_{4,\ell}(i+1)
              + \Big(y_{i,1} - y_{i,2} - 3y_{i,3} + \sum_{r=1}^{\ell}(y_{i,4,r} - y_{i,5,r} + y_{i,6,r})\Big) N_{4,\ell}(i) \nonumber \\
             &\quad + (-y_{i,2} - 2y_{i,3}) N_{4,\ell}(i-\ell-1)
              + \sum_{r=1}^{\ell}\Bigg\{ y_{i,5,r}\big[N_{4,\ell}(i+1) - N_{4,\ell}(i-\ell-1)\big] \nonumber \\
             &\quad + \sum_{j=i+r-\ell-1}^{i-1} \, \Big[(y_{i,4,r} + y_{i,5,r} - 2y_{i,6,r})N_{4,\ell}(j+1)
              + (-3y_{i,4,r} - 3y_{i,5,r} + 4y_{i,6,r})N_{4,\ell}(j) \nonumber \\
             &\qquad + (-2y_{i,4,r} - 2y_{i,5,r} + 2y_{i,6,r})N_{4,\ell}(j-\ell-1)\Big] \Bigg\},
\end{align}
\begin{equation}\label{gen_rule_gf4_l}
g(\mathbf{c}) = \sum_{i=0}^{m-1}g_i(c_i).
\end{equation}
Observe that the variables in \eqref{eqn_gf4_rule} are given by $k_{i,1}=y_{i,2} + 2y_{i,3}$, $k_{i,2}=y_{i,1} - y_{i,2} - 3y_{i,3} + \sum_{r=1}^{\ell}(y_{i,4,r} - y_{i,5,r} + y_{i,6,r})$, $k_{i,3}=-y_{i,2} - 2y_{i,3}$, $k_{i,4,r}= y_{i,5,r}$, $k_{i,5,r}=y_{i,4,r} + y_{i,5,r} - 2y_{i,6,r}$, $k_{i,6,r}=-3y_{i,4,r} - 3y_{i,5,r} + 4y_{i,6,r}$, and $k_{i,7,r}=-2y_{i,4,r} - 2y_{i,5,r} + 2y_{i,6,r}$. Combining \eqref{symb_cont_gf4_l} and \eqref{gen_rule_gf4_l} completes the proof and gives the encoding-decoding rule.
\end{proof}
\end{theorem}

\begin{remark}
The encoding-decoding rule derived in Theorem~\ref{res_gf4_l_rule} explicitly assumes $\ell \geq 1$. This assumption is necessary solely to accommodate special cases $S_1$, $S_2$, $S_3$, $S_4$, and $S_5$, which handle omitting shorter codewords as a result of intermediate $\mathbf{0}$ runs of length $r-1$, where $r \in \{1, 2, \ldots, \ell\}$. Since $r \geq 1$, the required range for $r$ becomes invalid when $\ell = 0$. Consequently, these specific short runs cannot mathematically exist, and cases $S_1$ through $S_5$ become infeasible. Conversely, the typical cases ($T_1$, $T_2$, $T_3$) and the special case $S_6$ remain structurally identical regardless of whether $\ell = 0$ or $\ell \geq 1$. Therefore, while the overarching logic aligns, the specific boundary case where $\ell = 0$ here reduces to the specialized rule previously established in Theorem~\ref{res_gf4_0_rule}.
\end{remark}

\textit{Sixth, we develop the encoding and decoding algorithms.}
The encoding and decoding algorithms follow the same systematic procedure detailed in \cite{Hareedy2022LOCO}, where the encoding-decoding rule is used to map between messages and codewords. Having illustrated the procedure for the GF$(4)$ codes with $\ell = 0$ above via Example~\ref{ex_decode} and Example~\ref{ex_encode}, we omit the detailed algorithmic description for brevity.

\section{RES-LOCO Coding Scheme Over GF$(8)$}\label{sec_gf8}

In this section, we introduce the GF$(8)$ RES-LOCO coding scheme that prevents sneak paths of circumference $4$. The code here is defined over GF$(8) = \{0, 1, \alpha, \alpha^2, \alpha^3, \alpha^4, \alpha^5, \alpha^6\}$, and it forbids sneak-path (SP) patterns. We adopt the general methodology in \cite{Hareedy2022LOCO} for the construction of the RES-LOCO code. The process begins with an examination of the forbidden patterns, using which the cardinality equations are derived, followed by the derivation of the encoding-decoding rule.

The GF$(8)$ RES-LOCO coding scheme adopts the following $\text{GF}(8) \longleftrightarrow [\text{GF}(2)]^3$ mapping-demapping:
\begin{align}\label{eqn_gf8map}
0 &\longleftrightarrow [0 ~ 0 ~ 0]^\mathrm{T}, \hspace{+2em} 1 \longleftrightarrow [0 ~ 0 ~ 1]^\mathrm{T}, \nonumber \\
\alpha &\longleftrightarrow [0 ~ 1 ~ 0]^\mathrm{T}, \hspace{+1.4em} \alpha^2 \longleftrightarrow [0 ~ 1 ~ 1]^\mathrm{T}, \nonumber \\
\alpha^3 &\longleftrightarrow [1 ~ 0 ~ 0]^\mathrm{T}, \hspace{+1.4em} \alpha^4 \longleftrightarrow [1 ~ 0 ~ 1]^\mathrm{T}, \\
\alpha^5 &\longleftrightarrow [1 ~ 1 ~ 0]^\mathrm{T}, \hspace{+1.4em} \alpha^6 \longleftrightarrow [1 ~ 1 ~ 1]^\mathrm{T}. \nonumber
\end{align}

Our forbidden pattern for preventing sneak paths of circumference $4$ is
\begin{align}\label{GF(8)_l_0_forbid}
\mathcal{T}^{8,0} = \{&1\alpha^2, 1\alpha^6, \alpha\alpha^2, \alpha\alpha^5, \alpha\alpha^6, \alpha^2 1, \alpha^2\alpha, \alpha^2\alpha^4,
\alpha^2\alpha^5, \alpha^2\alpha^6, \alpha^3\alpha^5, \alpha^3\alpha^6, \alpha^4\alpha^2, \alpha^4\alpha^5, \alpha^4\alpha^6, \nonumber \\
&\alpha^5\alpha, \alpha^5\alpha^2, \alpha^5\alpha^3, \alpha^5\alpha^4, \alpha^5\alpha^6, \alpha^6 1, \alpha^6\alpha,
\alpha^6\alpha^2, \alpha^6\alpha^3, \alpha^6\alpha^4, \alpha^6\alpha^5\}.
\end{align}
To further clarify, consider the forbidden pattern $1\alpha^6$. Here, we have the two columns $[0 ~ 0 ~ 1]^\mathrm{T}$ and $[1 ~ 1 ~ 1]^\mathrm{T}$ consecutively on the crossbar array, which results in a circumference-$4$ sneak path across the bottom two rows.

The formal definition of a GF$(8)$ RES-LOCO code is as follows:
\begin{definition}[GF$(8)$ RES-LOCO Code]\label{def:GF(8) RES-LOCO}
A RES-LOCO code, $\mathcal{RES}^{8,0}_{m}$, is defined by the following properties:
\begin{enumerate}
    \item Codewords in $\mathcal{RES}^{8,0}_{m}$ are defined over $\textup{GF}(8)$, the code alphabet, and are of length $m$ symbols.
    \item Codewords in $\mathcal{RES}^{8,0}_{m}$ are lexicographically ordered.
    \item Codewords in $\mathcal{RES}^{8,0}_{m}$ do not contain any patterns from the set $\mathcal{T}^{8,0}$.
    \item Any codeword satisfying the above properties is included in $\mathcal{RES}^{8,0}_{m}$.
\end{enumerate}
\end{definition}

Lexicographic ordering means codewords are ordered following the notion $0 < 1 < \alpha < \alpha^2 < \alpha^3 < \alpha^4 < \alpha^5 < \alpha^6$ and symbol significance reduces from left to right within the codeword.

\textit{First, we specify the group structure.} Groups of $\mathcal{RES}^{8,0}_{m}$ code are:
\begin{itemize}
    \item Group~$1$ contains all the codewords starting with 
          $0x$, where $x$ can be any of the eight symbols.
    \item Group~$2$ contains all the codewords starting with 
          $1\delta_1$, where $\delta_1 \in \{0, 1, \alpha, \alpha^3, \alpha^4, \alpha^5\}$.
    \item Group~$3$ contains all the codewords starting with 
          $\alpha\delta_2$, where $\delta_2 \in \{0, 1, \alpha, \alpha^3, \alpha^4\}$.
    \item Group~$4$ contains all the codewords starting with 
          $\alpha^2\delta_3$, where $\delta_3 \in \{0, \alpha^2, \alpha^3\}$.
    \item Group~$5$ contains all the codewords starting with 
          $\alpha^3\delta_4$, where $\delta_4 \in \{0, 1, \alpha, \alpha^2, \alpha^3, \alpha^4\}$.
    \item Group~$6$ contains all the codewords starting with 
          $\alpha^4\delta_5$, where $\delta_5 \in \{0, 1, \alpha, \alpha^3, \alpha^4\}$.
    \item Group~$7$ contains all the codewords starting with 
          $\alpha^5\delta_6$, where $\delta_6 \in \{0, 1, \alpha^5\}$.
    \item Group~$8$ contains all the codewords starting with 
          $\alpha^6\delta_7$, where $\delta_7 \in \{0, \alpha^6\}$.
\end{itemize}

\textit{Second, we enumerate the codewords.}
Let $N_{8,0}(m)$ denote the number of $\mathcal{RES}^{8,0}_{m}$ codewords of length $m$. Additionally, let $N_{8,0,i}(m)$ denote the number of $\mathcal{RES}^{8,0}_{m}$ codewords of length $m$ belonging to Group~$i$, where $i\in\{1, 2, 3, 4, 5, 6, 7, 8\}$. Thus, it follows that $N_{8,0}(m) = \sum_{i = 1}^{8}N_{8,0,i}(m)$.

From the GF$(8)$ mapping, we observe a symmetric structure. We map $1$ as $[0~0~1]^{\mathrm{T}}$ and $\alpha^3$ as $[1~0~0]^{\mathrm{T}}$, and binary columns with $1$ at the top or $1$ at the bottom result in the symmetry of Groups~$2$ and $5$. Similarly, $[0~1~1]^{\mathrm{T}}$ and $[1~1~0]^{\mathrm{T}}$ are binary columns with $0$ at the top or $0$ at the bottom, and thus Groups~$4$ and $7$ are symmetric. By definition of the group structure, Groups~$3$ and $6$ are of equal cardinalities. Consequently, the following cardinality relations hold:
\begin{align}
N_{8,0,2}(m) &= N_{8,0,5}(m), \label{eq:8_sym1} \\
N_{8,0,3}(m) &= N_{8,0,6}(m), \label{eq:8_sym2} \\
N_{8,0,4}(m) &= N_{8,0,7}(m). \label{eq:8_sym3}
\end{align}

\begin{theorem}\label{thm:N(m)_for_GF8_l0}
The cardinality $N_{8,0}(m)$ of a RES-LOCO code $\mathcal{RES}^{8,0}_{m}$, for $m\geq 5$, is given by the following recurrence relation:
\begin{equation}\label{eq:N(m)_for_GF8_l0}
N_{8,0}(m) = 7N_{8,0}(m-1) - 7N_{8,0}(m-2) - 10N_{8,0}(m-3) + 10N_{8,0}(m-4) + 2N_{8,0}(m-5),
\end{equation}
where the defined cardinalities are $N_{8,0}(0)\triangleq1$, $N_{8,0}(-1)\triangleq1$, $N_{8,0}(-2)\triangleq-1$, $N_{8,0}(-3)\triangleq4.5$, and $N_{8,0}(-4)\triangleq-23.5$.

\begin{proof}
    From the group structure, we observe that:
    \begin{equation}\label{thm3,eq1}
        N_{8,0,1}(m) = N_{8,0}(m-1),
    \end{equation}
    since any codeword of length $m-1$ can be concatenated to the right of symbol $0$.
    
    For Group~$2$, $\delta_1$ cannot be $\alpha^2$ and $\alpha^6$. Therefore, and by symmetry:
\begin{equation}\label{thm3,eq2}
    N_{8,0,2}(m) = N_{8,0,5}(m) = N_{8,0}(m-1) - N_{8,0,4}(m-1) - N_{8,0,8}(m-1).
\end{equation}   
    For Groups~$3$ and $6$, since $\delta_2$ (or $\delta_5$) cannot be $\alpha^2, \alpha^5, \alpha^6$, we have:
\begin{equation}\label{thm3,eq3}
    N_{8,0,3}(m) = N_{8,0,6}(m) = N_{8,0}(m-1) - N_{8,0,4}(m-1) - N_{8,0,7}(m-1) - N_{8,0,8}(m-1).
\end{equation}
    Using symmetry $N_{8,0,4}(m) = N_{8,0,7}(m)$, this simplifies to:
    \begin{equation}\label{thm3,eq4}
        N_{8,0,3}(m) = N_{8,0}(m-1) - 2N_{8,0,4}(m-1) - N_{8,0,8}(m-1).
    \end{equation}

    Now, we combine Groups~$4$ and $7$, resulting in the addition of all codewords starting with $\{0,\alpha^2,\alpha^3\} \cup \{0,1,\alpha^5\}$ from the left at length $m-1$, where each codeword starting with a $0$ is added twice. Therefore, and using symmetry: 
\begin{equation}\label{thm3,eq5}
    2N_{8,0,4}(m) = N_{8,0}(m-1) + N_{8,0,1}(m-1) - N_{8,0,3}(m-1) - N_{8,0,6}(m-1) - N_{8,0,8}(m-1).
\end{equation}
Using \eqref{thm3,eq1} and substituting $N_{8,0,3}(m)$ from \eqref{thm3,eq4} (using index $m-1$ on the right-hand side), we obtain:
\begin{align}
    2N_{8,0,4}(m) &= N_{8,0}(m-1) + N_{8,0}(m-2) \nonumber \\
    &\quad - 2\Big[N_{8,0}(m-2) - 2N_{8,0,4}(m-2) - N_{8,0,8}(m-2)\Big] \nonumber \\
    &\quad - N_{8,0,8}(m-1). \label{thm3,eq6}
\end{align}
Expanding the terms in \eqref{thm3,eq6} yields the following recurrence for $N_{8,0,4}$:
    \begin{equation}\label{thm3,eq9}
        2N_{8,0,4}(m) = N_{8,0}(m-1) - N_{8,0}(m-2) + 4N_{8,0,4}(m-2) + 2N_{8,0,8}(m-2) - N_{8,0,8}(m-1).
    \end{equation}
    
    For Group~$8$, since $\delta_7 \in \{0, \alpha^6\}$:
\begin{equation}\label{thm3,eq8}
N_{8,0,8}(m) = N_{8,0,1}(m-1) + N_{8,0,8}(m-1) = N_{8,0}(m-2) + N_{8,0,8}(m-1).
\end{equation}
    Rearranging \eqref{thm3,eq8} gives the difference relation:
    \begin{equation}\label{eq:N8_diff_rel}
        N_{8,0,8}(m) - N_{8,0,8}(m-1) = N_{8,0}(m-2).
    \end{equation}
    
    Now, we sum all group cardinalities to reach the recursion of the total cardinality $N_{8,0}(m)$. Using \eqref{thm3,eq1}, \eqref{thm3,eq2}, and \eqref{thm3,eq4}:
\begin{align}
    N_{8,0}(m) &= N_{8,0,1}(m) + 2N_{8,0,2}(m) + 2N_{8,0,3}(m) + 2N_{8,0,4}(m) + N_{8,0,8}(m) \nonumber \\ &= 5N_{8,0}(m-1) + \Big[2N_{8,0,4}(m) - 6N_{8,0,4}(m-1)\Big] + \Big[N_{8,0,8}(m) - 4N_{8,0,8}(m-1)\Big]. \label{thm3,eq10}
\end{align}
    We eliminate $N_{8,0,4}(\cdot)$ by computing $N_{8,0}(m) - 2N_{8,0}(m-2)$ from \eqref{thm3,eq10}. This operation groups the $N_{8,0,4}(\cdot)$ terms into the form $2N_{8,0,4}(k) - 4N_{8,0,4}(k-2)$, which can be directly replaced using \eqref{thm3,eq9} by $N_{8,0}(k-1) - N_{8,0}(k-2) + 2N_{8,0,8}(k-2) - N_{8,0,8}(k-1)$.
    Substituting these relations and simplifying the resulting $N_{8,0,8}(\cdot)$ terms using \eqref{eq:N8_diff_rel} yields Equation~\eqref{eq:intermediate_N8}, which depends only on $N_{8,0}(\cdot)$ and $N_{8,0,8}(\cdot)$:
\begin{align}
    N_{8,0}(m) &= 6N_{8,0}(m-1) - 2N_{8,0}(m-2) - 8N_{8,0}(m-3) + 3N_{8,0}(m-4) \nonumber \\
    &\quad + N_{8,0,8}(m) - 4N_{8,0,8}(m-1) - N_{8,0,8}(m-2) + 5N_{8,0,8}(m-3). \label{eq:intermediate_N8}
\end{align}
    To eliminate the remaining $N_{8,0,8}(\cdot)$ terms, we compute the difference $N_{8,0}(m) - N_{8,0}(m-1)$. This operation changes the $N_{8,0,8}(\cdot)$ terms in \eqref{eq:intermediate_N8} into differences of the form $N_{8,0,8}(k) - N_{8,0,8}(k-1)$. Applying \eqref{eq:N8_diff_rel} replaces each such difference with $N_{8,0}(k-2)$, resulting in an equation that depends solely on $N_{8,0}(\cdot)$. Rearranging the terms yields the final recurrence in \eqref{eq:N(m)_for_GF8_l0}, which completes the proof.

We note that the defined cardinalities are calculated from the ``known'' cardinalities. In particular, we use the cardinalities $N_{8,0}(1)$, $N_{8,0}(2)$, $N_{8,0}(3)$, $N_{8,0}(4)$, and $N_{8,0}(5)$ to build five equations in five unknowns $N_{8,0}(0)$, $N_{8,0}(-1)$, $N_{8,0}(-2)$, $N_{8,0}(-3)$, and $N_{8,0}(-4)$. Then, we solve these equations. Alternatively, we can also build five equations in the same five unknowns by observing that $N_{8,0,1}(1) = N_{8,0,2}(1) = N_{8,0,3}(1) = N_{8,0,4}(1) = N_{8,0,8}(1) = 1$, and the recursion of these group cardinalities is explicitly found in Corollary~\ref{cor:subgroup_cardinalities}.
\end{proof}
\end{theorem}

\begin{corollary}\label{cor:subgroup_cardinalities}
The individual group cardinalities can be expressed solely in terms of the total cardinality $N_{8,0}(\cdot)$:
\begin{align}
N_{8,0,1}(m) &= N_{8,0}(m-1), \label{eq:N1_formula} \\
N_{8,0,2}(m) &= \frac{1}{14}[17N_{8,0}(m-1) - 19N_{8,0}(m-2) - 36N_{8,0}(m-3) + 6N_{8,0}(m-4) + 2N_{8,0}(m-5)], \label{eq:N2_formula} \\
N_{8,0,3}(m) &= \frac{1}{7}[17N_{8,0}(m-1) - 61N_{8,0}(m-2) - 22N_{8,0}(m-3) + 90N_{8,0}(m-4) + 16N_{8,0}(m-5)], \label{eq:N3_formula} \\
N_{8,0,4}(m) &= \frac{1}{14}[-16N_{8,0}(m-1) + 127N_{8,0}(m-2) - 4N_{8,0}(m-3) - 200N_{8,0}(m-4) - 34N_{8,0}(m-5)], \label{eq:N4_formula} \\
N_{8,0,8}(m) &= N_{8,0}(m-1) - 5N_{8,0}(m-2) + 2N_{8,0}(m-3) + 12N_{8,0}(m-4) + 2N_{8,0}(m-5). \label{eq:N8_formula}
\end{align}
\begin{proof}
The derivation proceeds by solving equations for each group sequentially.

\noindent\textbf{1. Derivation of $N_{8,0,8}(m)$:}
We determine $N_{8,0,8}(m)$ by equating the right-hand side of \eqref{eq:intermediate_N8} to that of \eqref{eq:N(m)_for_GF8_l0}.
First, we simplify the $N_{8,0,8}(\cdot)$ terms in \eqref{eq:intermediate_N8}. In particular, by repeatedly applying the difference relation from \eqref{eq:N8_diff_rel}, the $N_{8,0,8}(\cdot)$ terms can be simplified as follows:
\begin{equation}\label{eq:N8_reduction_step}
N_{8,0,8}(m) - 4N_{8,0,8}(m-1) - N_{8,0,8}(m-2) + 5N_{8,0,8}(m-3) = N_{8,0}(m-2) - 3N_{8,0}(m-3) - 4N_{8,0}(m-4) + N_{8,0,8}(m-3).
\end{equation}
Substituting~\eqref{eq:N8_reduction_step} back into \eqref{eq:intermediate_N8} yields:
\begin{equation}\label{eq:N_with_N8_isolated}
N_{8,0}(m) = 6N_{8,0}(m-1) - N_{8,0}(m-2) - 11N_{8,0}(m-3) - N_{8,0}(m-4) + N_{8,0,8}(m-3).
\end{equation}
We now equate the right-hand side of \eqref{eq:N_with_N8_isolated} to the expansion of $N_{8,0}(m)$ provided by \eqref{eq:N(m)_for_GF8_l0}, which is the main recursion:
\begin{align*}
&6N_{8,0}(m-1) - N_{8,0}(m-2) - 11N_{8,0}(m-3) - N_{8,0}(m-4) + N_{8,0,8}(m-3) \\
&= 7N_{8,0}(m-1) - 7N_{8,0}(m-2) - 10N_{8,0}(m-3) + 10N_{8,0}(m-4) + 2N_{8,0}(m-5).
\end{align*}
Solving for $N_{8,0,8}(m-3)$ gives:
\begin{equation}\label{eq:N8_shifted_result}
N_{8,0,8}(m-3) = N_{8,0}(m-1) - 6N_{8,0}(m-2) + N_{8,0}(m-3) + 11N_{8,0}(m-4) + 2N_{8,0}(m-5).
\end{equation}
Shifting the indices such that $m \to m+3$ and substituting the future terms $N_{8,0}(m+2)$ and $N_{8,0}(m+1)$ using \eqref{eq:N(m)_for_GF8_l0} results in the recursive expression in \eqref{eq:N8_formula}.

\noindent\textbf{2. Derivation of $N_{8,0,4}(m)$:}
We begin by substituting the formula for $N_{8,0,8}(m)$ from \eqref{eq:N8_formula} into the relation derived in \eqref{thm3,eq10}. This yields the following recurrence for $N_{8,0,4}(\cdot)$ in terms of $N_{8,0}(\cdot)$:
\begin{equation}\label{eq:N4_step1}
2N_{8,0,4}(m) - 6N_{8,0,4}(m-1) = 5N_{8,0}(m-1) - 26N_{8,0}(m-2) - 4N_{8,0}(m-3) + 46N_{8,0}(m-4) + 8N_{8,0}(m-5).
\end{equation}
To replace the $N_{8,0,4}(m-1)$ term, we consider the shifted version of \eqref{eq:N4_step1} computed at index $m-1$ and multiplied by 3. We then compute the linear combination:
\[
\Big[2N_{8,0,4}(m) - 6N_{8,0,4}(m-1)\Big] + 3\Big[2N_{8,0,4}(m-1) - 6N_{8,0,4}(m-2)\Big].
\]
This simplifies the left-hand side to $2N_{8,0,4}(m) - 18N_{8,0,4}(m-2)$.
By applying this operation to the right-hand side of \eqref{eq:N4_step1}, i.e., RHS$(m)+3$ RHS$(m-1)$, we arrive at the refined equation:
\begin{align}\label{eq:N4_step3}
2N_{8,0,4}(m) - 18N_{8,0,4}(m-2) &= 17N_{8,0}(m-1) - 95N_{8,0}(m-2) + 2N_{8,0}(m-3) \nonumber \\
&\quad + 154N_{8,0}(m-4) + 26N_{8,0}(m-5).
\end{align}
We then subtract \eqref{eq:N4_step3} from \eqref{thm3,eq9}, substitute $N_{8,0,8}(\cdot)$ from \eqref{eq:N8_formula}, and replace $m$ with $m+2$ to reach the final closed-form expression presented in \eqref{eq:N4_formula}.

\noindent\textbf{3. Derivation of $N_{8,0,2}(m)$:}
We compute $N_{8,0,2}(m)$ directly from \eqref{thm3,eq2}: 
\[
N_{8,0,2}(m) = N_{8,0}(m-1) - N_{8,0,4}(m-1) - N_{8,0,8}(m-1).
\]
Substituting the index-shifted version of expressions \eqref{eq:N8_formula} and \eqref{eq:N4_formula} into this relation and summing the coefficients for each term $N_{8,0}(m-k)$ yields the result in \eqref{eq:N2_formula}.

\noindent\textbf{4. Derivation of $N_{8,0,3}(m)$:}
We derive $N_{8,0,3}(m)$ using \eqref{thm3,eq4}: 
\[
N_{8,0,3}(m) = N_{8,0}(m-1) - 2N_{8,0,4}(m-1) - N_{8,0,8}(m-1).
\]
Substituting the index-shifted version of expressions \eqref{eq:N8_formula} and \eqref{eq:N4_formula} into this relation and summing the coefficients for each term $N_{8,0}(m-k)$ yields the result in \eqref{eq:N3_formula}.
\end{proof}
\end{corollary}
\textit{Third and Fourth, we find the special cases and symbol contribution.}
After calculating the necessary cardinality relations, we advance to the next stage, which is to define the typical and special cases to be used when calculating the contributions of symbols. We define the contribution of a given symbol $c_i$ to the overall codeword index $g(\mathbf{c})$ as the number of smaller-length codewords that we can generate by replacing that symbol with $c'_i < c_i$ and the symbols to its right with valid constrained alternatives. Valid here means we can concatenate them from the right to $c_{m-1} c_{m-2} \dots c_{i+2} c_{i+1}$, the symbols of $\mathbf{c}$, without violating the constraint. To calculate the contribution of symbol $c_i$, we must look at the preceding symbols $c_{m-1} \dots c_{i+2} c_{i+1}$. This is necessary because replacing $c_i c_{i-1} \dots c_0$ with shorter-length codewords might create a forbidden pattern once concatenated to $c_{m-1} \dots c_{i+2} c_{i+1}$. If this concatenation creates a forbidden pattern, which we characterize as a ``special'' case, some of the shorter-length codewords need to be omitted to correctly calculate the contribution of the symbol at hand. Otherwise, all shorter-length codewords can be concatenated and counted in the symbol contribution, which we characterize as a ``typical'' case. We then find the symbol contribution for each case.

\begin{theorem}\label{res_gf8_0_rule}
The encoding-decoding rule of a RES-LOCO code $\mathcal{RES}^{8,0}_{m}$, for $m\geq2$, is given by:
\begin{equation}\label{eqn_gf8_0_rule}
g(\mathbf{c}) = \sum_{i=0}^{m-1} \big[ w_{i,1} N_{8,0}(i) + w_{i,2} N_{8,0}(i-1) + w_{i,3} N_{8,0}(i-2) + w_{i,4} N_{8,0}(i-3) + w_{i,5} N_{8,0}(i-4) \big],
\end{equation}
where $w_{i,1}$, $w_{i,2}$, $w_{i,3}$, $w_{i,4}$, and $w_{i,5}$ are defined according to the special/typical case of symbol $c_i$ in $\mathbf{c}$. In particular, $[w_{i,1} ~ w_{i,2} ~ w_{i,3} ~ w_{i,4} ~ w_{i,5}] = \frac{1}{14} \mathbf{Y}^{\mathrm{T}} \mathbf{C}$, where the vector $\mathbf{Y}$ and the matrix $\mathbf{C}$ are given in \eqref{eq:merge_flags_gf8} and \eqref{eqn_gf8_C}, respectively.

\begin{proof}
Using the forbidden patterns that we have previously defined, we identify the typical and special cases as well as determine their corresponding symbol contributions $g_i(c_i)$ as follows.

\textit{Typical cases:}
\begin{itemize}
\item Typical case $T_1$, $c_{i+1}c_i=\delta_1 1$, where $\delta_1 \in \text{GF}(8) \setminus \{\alpha^2, \alpha^6\}$:
\begin{equation}
g_i(c_i) = N_{8,0,1}(i+1)
\end{equation}
since $c_i = 1$ enables only the smaller symbol $c'_i = 0$, contributing the codewords of Group~$1$.

\item Typical case $T_2$, $c_{i+1}c_i=\delta_1 \alpha$, where $\delta_1 \in \text{GF}(8) \setminus \{\alpha^2, \alpha^5, \alpha^6\}$:
\begin{equation}
g_i(c_i) = N_{8,0,1}(i+1) + N_{8,0,2}(i+1)
\end{equation}
since $c_i = \alpha$ enables its smaller symbols $c'_i \in \{0, 1\}$, contributing the codewords of Groups~$1$ and~$2$.

\item Typical case $T_3$, $c_{i+1}c_i=\delta_2 \alpha^2$, where $\delta_2 \in \{0, \alpha^3\}$:
\begin{equation}
g_i(c_i) = N_{8,0,1}(i+1) + N_{8,0,2}(i+1) + N_{8,0,3}(i+1)
\end{equation}
since $c_i = \alpha^2$ enables its smaller symbols, and the preceding symbol being in $\{0, \alpha^3\}$ creates no forbidden pattern. Thus, the codewords of Groups~$1$ through~$3$ all contribute.

\item Typical case $T_4$, $c_{i+1}c_i=\delta_2 \alpha^3$, where $\delta_2 \in \{0, \alpha^3\}$:
\begin{equation}
g_i(c_i) = N_{8,0,1}(i+1) + N_{8,0,2}(i+1) + N_{8,0,3}(i+1) + N_{8,0,4}(i+1)
\end{equation}
since $c_i = \alpha^3$ enables all its smaller symbols without any restriction resulting from the preceding symbol. Thus, the codewords of Groups~$1$ through~$4$ all contribute.

\item Typical case $T_5$, $c_{i+1}c_i=\delta_2 \alpha^4$, where $\delta_2 \in \{0, \alpha^3\}$:
\begin{equation}
g_i(c_i) = N_{8,0,1}(i+1) + N_{8,0,2}(i+1) + N_{8,0,3}(i+1) + N_{8,0,4}(i+1) + N_{8,0,5}(i+1)\vspace{-0.1em}
\end{equation}
since $c_i = \alpha^4$ enables all its smaller symbols without restriction, and thus the codewords of Groups~$1$ through~$5$ all contribute.

\item Typical case $T_6$, $c_{i+1}c_i=0\alpha^5$:
\begin{equation}
g_i(c_i) = N_{8,0,1}(i+1) + N_{8,0,2}(i+1) + N_{8,0,3}(i+1) + N_{8,0,4}(i+1) + N_{8,0,5}(i+1) + N_{8,0,6}(i+1)
\end{equation}
since $c_i = \alpha^5$ enables all its smaller symbols, and the preceding $0$ initiates no forbidden pattern, and thus the codewords of Groups~$1$ through~$6$ all contribute.

\item Typical case $T_7$, $c_{i+1}c_i=0\alpha^6$:
\begin{equation}
g_i(c_i) = N_{8,0,1}(i+1) + N_{8,0,2}(i+1) + N_{8,0,3}(i+1) + N_{8,0,4}(i+1) + N_{8,0,5}(i+1) + N_{8,0,6}(i+1) + N_{8,0,7}(i+1)
\end{equation}
since $c_i = \alpha^6$ enables all its smaller symbols, and the preceding $0$ initiates no forbidden pattern, and thus the codewords of all seven relevant groups contribute.
\end{itemize}

\textit{Special cases:}
\begin{itemize}
\item Special case $S_1$, $c_{i+1}c_i=\alpha^2\alpha^2$:
\begin{equation}
g_i(c_i) = N_{8,0,1}(i+1)
\end{equation}
since the preceding $\alpha^2$ disables all but Group~$1$ codewords when $c_i = \alpha^2$ is replaced by a smaller symbol.

\item Special case $S_2$, $c_{i+1}c_i=\delta_3 \alpha^3$, where $\delta_3 \in \{1, \alpha, \alpha^4\}$:
\begin{equation}
g_i(c_i) = N_{8,0,1}(i+1) + N_{8,0,2}(i+1) + N_{8,0,3}(i+1)
\end{equation}
since the preceding symbol in $\{1, \alpha, \alpha^4\}$ enables codewords in all relevant groups except Group~$4$ relative to the typical contribution of $c_i = \alpha^3$.

\item Special case $S_3$, $c_{i+1}c_i=\alpha^2\alpha^3$:
\begin{equation}
g_i(c_i) = N_{8,0,1}(i+1) + N_{8,0,4}(i+1)
\end{equation}
since the preceding $\alpha^2$ disables codewords in all relevant groups other than Groups~$1$ and~$4$.

\item Special case $S_4$, $c_{i+1}c_i=\delta_3 \alpha^4$, where $\delta_3 \in \{1, \alpha, \alpha^4\}$:
\begin{equation}
g_i(c_i) = N_{8,0,1}(i+1) + N_{8,0,2}(i+1) + N_{8,0,3}(i+1) + N_{8,0,5}(i+1)
\end{equation}
since the preceding symbol in $\{1, \alpha, \alpha^4\}$ enables codewords in all relevant groups except Group~$4$ relative to the typical contribution of $c_i = \alpha^4$.

\item Special case $S_5$, $c_{i+1}c_i=1\alpha^5$:
\begin{equation}
g_i(c_i) = N_{8,0,1}(i+1) + N_{8,0,2}(i+1) + N_{8,0,3}(i+1) + N_{8,0,5}(i+1) + N_{8,0,6}(i+1)
\end{equation}
since the preceding $1$ enables codewords in all relevant groups except Group~$4$ relative to the typical contribution of $c_i = \alpha^5$.

\item Special case $S_6$, $c_{i+1}c_i=\alpha^5\alpha^5$:
\begin{equation}
g_i(c_i) = N_{8,0,1}(i+1) + N_{8,0,2}(i+1)
\end{equation}
since the preceding $\alpha^5$ disables codewords in all relevant groups other than Groups~$1$ and~$2$ when $c_i = \alpha^5$ is replaced by a smaller symbol.

\item Special case $S_7$, $c_{i+1}c_i=\alpha^6\alpha^6$:
\begin{equation}
g_i(c_i) = N_{8,0,1}(i+1)
\end{equation}
since the preceding $\alpha^6$ disables codewords in all relevant groups other than Group~$1$ when $c_i = \alpha^6$ is replaced by a smaller symbol according to the lexicographic ordering.
\end{itemize}

If $c_i$ corresponds to the leftmost symbol of the codeword, i.e., $c_i = c_{m-1}$, then we assume that $c_{i+1} = c_m \triangleq 0$ for the purposes of identifying typical and special cases.

\textit{Fifth, we formulate the encoding-decoding rule.}
After outlining the symbol contributions for each special and typical case to the overall codeword index, we are now able to determine the merged indexing rule, which combines all the aforementioned cases. We define indicator functions $y_{i,T_k}$ for $k \in \{1,2,\ldots,7\}$ and $y_{i,S_k}$ for $k \in \{1,2,\ldots,7\}$ that attain the value $1$ whenever the relevant case is the corresponding case, and have the value $0$ otherwise. To simplify the analysis, we define merging flags that combine indicator functions for cases with identical contributions as follows:
\begin{equation}\label{eq:merge_flags_gf8}
\begin{matrix}
Y_{i,1} = y_{i,T_1}+y_{i,S_1}+y_{i,S_7}, & Y_{i,2} = y_{i,T_2}+y_{i,S_6}, \\
Y_{i,3} = y_{i,T_3} + y_{i,S_2}, & Y_{i,4} = y_{i,T_4}, \\
Y_{i,5} = y_{i,T_5}, & Y_{i,6} = y_{i,T_6}, \\
Y_{i,7} = y_{i,T_7}, & Y_{i,8} = y_{i,S_5}, \\
Y_{i,9} = y_{i,S_4}, & Y_{i,10} = y_{i,S_3}.
\end{matrix}
\end{equation}
Using the symmetry relations $N_{8,0,2}(\cdot)=N_{8,0,5}(\cdot)$, $N_{8,0,3}(\cdot)=N_{8,0,6}(\cdot)$, and $N_{8,0,4}(\cdot)=N_{8,0,7}(\cdot)$, the indexing rule becomes:
\begin{align}\label{symb_cont_gf8_0}
g_i(c_i) &= \sum_{k=1}^{10} Y_{i,k} N_{8,0,1}(i+1)
 + \left(\sum_{k=2}^{4} Y_{i,k} + 2\sum_{k=5}^{9} Y_{i,k}\right) N_{8,0,2}(i+1) \nonumber \\
&\quad + \left(\sum_{k=3}^{5} Y_{i,k} + 2\sum_{k=6}^{8} Y_{i,k} + Y_{i,9}\right) N_{8,0,3}(i+1)
 + \left(\sum_{k=4}^{6} Y_{i,k} + 2Y_{i,7} + Y_{i,10}\right) N_{8,0,4}(i+1).
\end{align}

Using the group cardinality relations we derived in Corollary~\ref{cor:subgroup_cardinalities}, this can be expressed in matrix form as:
\begin{equation}
g_i(c_i) = \frac{1}{14} \mathbf{Y}^{\mathrm{T}} \mathbf{C} \mathbf{N},
\end{equation}
where $\mathbf{Y} = [Y_{i,1} ~ Y_{i,2} ~ \cdots ~ Y_{i,10}]^{\mathrm{T}}$, $\mathbf{N} = [N_{8,0}(i) ~ N_{8,0}(i-1) ~ \cdots ~ N_{8,0}(i-4)]^{\mathrm{T}}$, and
\begin{equation}\label{eqn_gf8_C}
\mathbf{C} = \begin{bmatrix}
14 & 0 & 0 & 0 & 0 \\
31 & -19 & -36 & 6 & 2 \\
65 & -141 & -80 & 186 & 34 \\
49 & -14 & -84 & -14 & 0 \\
66 & -33 & -120 & -8 & 2 \\
100 & -155 & -164 & 172 & 34 \\
84 & -28 & -168 & -28 & 0 \\
116 & -282 & -160 & 372 & 68 \\
82 & -160 & -116 & 192 & 36 \\
-2 & 127 & -4 & -200 & -34
\end{bmatrix}.
\end{equation}
Observe that the coefficients in \eqref{eqn_gf8_0_rule} are the elements of the row vector $[w_{i,1} ~ w_{i,2} ~ w_{i,3} ~ w_{i,4} ~ w_{i,5}] = \frac{1}{14} \mathbf{Y}^{\mathrm{T}} \mathbf{C}$. Combining the above with
\begin{equation}
g(\mathbf{c}) = \sum_{i=0}^{m-1}g_i(c_i)
\end{equation}
completes the proof and gives the encoding-decoding rule.
\end{proof}
\end{theorem}

\begin{remark}
Starting from \eqref{symb_cont_gf8_0}, there are two approaches for determining the individual group cardinalities. The first approach applies Corollary~\ref{cor:subgroup_cardinalities} to express each individual cardinality in terms of the total cardinalities, $N_{8,0}(\cdot)$, at various lengths. The second approach uses the fact that individual group cardinalities satisfy the recurrence relation given in Theorem~\ref{thm:N(m)_for_GF8_l0} of the total cardinality, i.e., the growth rate of all group cardinalities is identical to that of the total one (see Remark~\ref{rmk_cayley}). This approach bypasses the total cardinality entirely, requiring only the general recursion equation (in Theorem~\ref{thm:N(m)_for_GF8_l0}) and the initial cardinalities for each specific group. Because the system is deterministic and observable, these initial values are always unique. While this second approach offers lower computational complexity, it introduces a higher storage overhead. In this paper, we proceed with the first approach.
\end{remark}

\begin{remark}\label{rmk_cayley}
The reason why individual group cardinalities must satisfy the recursive relation of the total cardinality as stated above is an application of the Cayley-Hamilton theorem. Consider GF$(4)$ RES-LOCO codes with $\ell=0$. For this case, the characteristic-polynomial term of which the maximum real positive eigenvalue $\lambda_{\max}$ is a root gives:
\begin{equation}\label{char_gf4_l0}
\lambda^4_{\max} - 4 \lambda^3_{\max} + 2 \lambda^2_{\max} + 2 \lambda_{\max} = 0.
\end{equation}
Using the Cayley-Hamilton theorem, the transition matrix $\mathbf{A}$, which is the adjacency matrix of the transition diagram describing the code, also satisfies the same equation \eqref{char_gf4_l0}, leading to:
\begin{equation}\label{cayley_ham}
\mathbf{A}^m = 4 \mathbf{A}^{m-1} - 2 \mathbf{A}^{m-2} - 2 \mathbf{A}^{m-3}.
\end{equation}
Observe that \eqref{cayley_ham} gives exactly the total cardinality recursion if we multiply both sides from the right by the column vector $[1 ~ 1 ~ \dots ~ 1 ]^\mathrm{T}$ and from the left by the row vector $[1 ~ 1 ~ \dots ~ 1 ]$. Now, if we multiply both sides of \eqref{cayley_ham} only from the right by the column vector $[1 ~ 1 ~ \dots ~ 1 ]^\mathrm{T}$, different rows of the resulting vectors give different group cardinalities at different lengths. Since the recursion remains exactly the same, group cardinalities maintain the same total-cardinality recursion.
\end{remark}

\textit{Sixth, we develop the encoding and decoding algorithms.}
The encoding and decoding algorithms follow the same systematic procedure detailed in \cite{Hareedy2022LOCO}, where the encoding-decoding rule is used to map between messages and codewords. Having established this rule for the GF$(8)$ codes with $\ell = 0$ above, we omit the algorithmic description. However, we illustrate the main ideas via the following example focused on decoding.

\begin{example}[Decoding]
We find the index of codeword $\alpha^6\alpha^6\alpha^6$ in $\mathcal{RES}_3^{8,0}$.

For $i=0$, we have $c_1c_0 = \alpha^6\alpha^6 \Rightarrow y_{S_7}=1 \Rightarrow Y_1=1$:
\begin{equation}
g_0(c_0) = \frac{1}{14}[14N_{8,0}(0)] = N_{8,0}(0) = 1.
\end{equation}

For $i=1$, we have $c_2c_1 = \alpha^6\alpha^6 \Rightarrow y_{S_7}=1 \Rightarrow Y_1=1$:
\begin{equation}
g_1(c_1) = \frac{1}{14}[14N_{8,0}(1)] = N_{8,0}(1) = 8.
\end{equation}

For $i=2$, we have $c_3c_2 = 0\alpha^6 \Rightarrow y_{T_7}=1 \Rightarrow Y_7=1$:
\begin{align}
g_2(c_2) &= \frac{1}{14}[84N_{8,0}(2) - 28N_{8,0}(1)
 - 168N_{8,0}(0) - 28N_{8,0}(-1)] \nonumber \\
&= 6N_{8,0}(2) - 2N_{8,0}(1)
 - 12N_{8,0}(0) - 2N_{8,0}(-1) \nonumber \\
&= 6(38) - 2(8) - 12 - 2 = 198.
\end{align}

Therefore, $g(\alpha^6\alpha^6\alpha^6) = 1 + 8 + 198 = 207$. This result is consistent with $N_{8,0}(3) = 208$, as $\alpha^6\alpha^6\alpha^6$ is the last codeword of length $3$.
\end{example}

\section{Bridging, Self-Clocking, and Capacity Achievability}\label{sec_rates}

In the previous sections, we designed a series of codes defined over \text{GF}($4$) and \text{GF}($8$), which we denote as RES-LOCO codes, that eliminate or reduce the occurrence of sneak paths in isolated segments of a 2D memristor crossbar array consisting of two and three consecutive rows, respectively. From here, the next step is to perform bridging and self-clocking to address the following two issues, respectively: preventing forbidden patterns (i.e., sneak paths of certain circumferences) from occurring when two or more codewords are concatenated and ensuring that at least one symbol transition takes place in every concatenation of a codeword and its bridging symbols. While the system objective of bridging is obvious as it directly stems from mitigating the sneak-path problem, self-clocking is needed to maintain self calibration by preventing the current reader from measuring values in the same range consecutively. From there, we show capacity achievability of RES-LOCO codes as well as finite-length rates as the codeword length grows.

Bridging refers to the addition of a certain number of symbols at the end of each codeword such that the possibility of the creation of new forbidden patterns that arise from the concatenation of two or more codewords is eliminated. We designate the number of symbols added to each codeword as $\eta_3$. In certain cases, there might be multiple combinations of $\eta_3$ symbols that can be placed in between two codewords that all result in no new forbidden patterns at the transitions. In these cases, an additional $\eta_2$ message bits can be coded within the bridging segment, depending on the number of these combinations, which increases the finite-length rate. Self-clocking requires that not all symbols of a codeword associated with its bridging symbols be the same, so that there is at least one symbol transition within the codeword and its bridging symbols. To ensure this, we try to optimize the bridging patterns to induce at least one symbol transition in cases where the pre-bridging RES-LOCO codeword is comprised of the same symbol repeated for the whole length, and remove a certain number of codewords from the codebook if necessary, whose amount we denote by $\eta_1$.

In the presence of bridging and self-clocking, we define the code rate for our RES-LOCO \text{GF}($4$), for a generic $\ell$, and \text{GF}($8$) codes as follows, where $m$ denotes the original codeword length:
\begin{equation}
    R_{\text{GF}(4)} = \frac{\lfloor \log_2(N_{4,\ell}(m) - \eta_1)\rfloor + \eta_2}{2(m + \eta_3)},
\end{equation}
\begin{equation}
    R_{\text{GF}(8)} = \frac{\lfloor \log_2(N_{8,0}(m) - \eta_1)\rfloor + \eta_2}{3(m + \eta_3)}.
\end{equation}
The following two subsections provide bridging scenarios, tables, and finite-length rates for our two RES-LOCO codes.

\subsection{RES-LOCO \text{GF}($4$) Bridging and Self-Clocking}

\text{GF}($4$) RES-LOCO codes prevent any and all sneak paths up to a horizontal dimension of $\ell+1$, with a vertical dimension of $1$ due to the grouping of two consecutive rows. To effectively combat against the case where one codeword ends with $\alpha^2$ and the other begins with $1$ or $\alpha$ (and vice versa), at least $\ell+1$ bridging symbols need to be inserted in between the codewords. As such, we investigate bridging in three cases: $\eta_3 = \ell + 1$, $\eta_3 = \ell + 2$, and $\eta_3 = \ell + 3$.

In the $\eta_3 = \ell + 1$ case, bridging with all zeros ($0^{\ell + 1}$) is necessary to accommodate the aforementioned case where one codeword ends with $\alpha^2$ and the other begins with $1$ or $\alpha$ (and vice versa), since inserting any symbol other than zero at any location within the bridging segment would create a sneak path of horizontal dimension smaller than or equal to $\ell+1$. Hence, we are unable to encode any extra bits through bridging (since only one option is available) and have to remove the all-zero RES-LOCO codeword ($0^{m}$) to ensure self-clocking. Therefore, $\eta_1 = 1$ and $\eta_2 = 0$.

In the $\eta_3 = \ell + 2$ case, we find that there are at least $4$ possible bridging patterns (the minimum is $4$) for every possible configuration of concatenated~codewords, which results in $\eta_2 = \lfloor \log_2(4) \rfloor = 2$ bits. Additionally, we chose to remove the all-zero  and all-$\alpha^2$ codewords, resulting in $\eta_1 = 2$. Table~\ref{table_GF_4_l_2} outlines the possible bridging patterns for $\eta_3 = \ell + 2$.
\begin{table}
\caption{Possible Bridging Patterns for \text{GF}($4$) RES-LOCO Codes With $\eta_3 = \ell + 2$, $p, q \geq 0$}
\label{table_GF_4_l_2}
\centering
\footnotesize 
\renewcommand{\arraystretch}{1.2} 
\begin{tabularx}{\columnwidth}{|c| >{\centering\arraybackslash}X |c|}
\hline
\thead{Left-hand LOCO \\ ending with} & \thead{Possible bridging patterns} & \thead{Right-hand LOCO \\ beginning with} \\
\hline
$\alpha^2\mathbf{0}^p$ & $\mathbf{0}^{\ell+2}$, $\mathbf{0}^{\ell+1}1$, $\mathbf{0}^{\ell+1}\alpha$, $\alpha^2\mathbf{0}^{\ell+1}$ & $\mathbf{0}^q1$, $\mathbf{0}^q\alpha$ \\
\hline
$1\mathbf{0}^p$, $\alpha\mathbf{0}^p$ & $\mathbf{0}^{\ell+2}$, $1\mathbf{0}^{\ell+1}$, $\alpha\mathbf{0}^{\ell+1}$, $\mathbf{0}^{\ell+1}\alpha^2$ & $\mathbf{0}^q\alpha^2$ \\
\hline
$\alpha^2\mathbf{0}^p$ & $\mathbf{0}^{\ell+2}$, $\alpha^2\mathbf{0}^{\ell+1}$, $0\alpha^2\mathbf{0}^{\ell}$, $\alpha^2\alpha^2\mathbf{0}^{\ell}$ & $\mathbf{0}^q\alpha^2$ \\
\hline
$1\mathbf{0}^p$, $\alpha\mathbf{0}^p$ & $\mathbf{0}^{\ell+2}$, $1\mathbf{0}^{\ell+1}$, $\mathbf{0}^{\ell+1}1$, $\mathbf{0}^{\ell+1}\alpha$ & $\mathbf{0}^q1$, $\mathbf{0}^q\alpha$ \\
\hline
\end{tabularx}
\end{table}

In the $\eta_3 = \ell + 3$ case, we find that there are at least $8$ possible bridging patterns (the minimum is $14$) for every possible configuration of concatenated codewords, which results in $\eta_2 = \lfloor \log_2(8) \rfloor = 3$ bits. Additionally, we chose to remove the all-zero and all-$\alpha^2$ codewords, resulting in $\eta_1 = 2$. Table~\ref{table_GF_4_l_3} outlines the possible bridging configurations for $\eta_3 = \ell + 3$.

\begin{table}
\caption{Possible Bridging Patterns for \text{GF}($4$) RES-LOCO Codes With $\eta_3 = \ell + 3$, $p, q \geq 0$}
\label{table_GF_4_l_3}
\centering
\footnotesize 
\renewcommand{\arraystretch}{1.2} 
\begin{tabularx}{\columnwidth}{|c| >{\centering\arraybackslash}X |c|}
\hline
\thead{Left-hand LOCO \\ ending with} & \thead{Possible bridging scenarios} & \thead{Right-hand LOCO \\ beginning with} \\
\hline
$\alpha^2\mathbf{0}^p$ & $\mathbf{0}^{\ell+2}1$, $\mathbf{0}^{\ell+2}\alpha$, $\mathbf{0}^{\ell+1}10$, $\mathbf{0}^{\ell+1}11$, $\mathbf{0}^{\ell+1}1\alpha$, $\mathbf{0}^{\ell+1}\alpha0$, $\mathbf{0}^{\ell+1}\alpha1$, $\mathbf{0}^{\ell+1}\alpha\alpha$ & $\mathbf{0}^q1$, $\mathbf{0}^q\alpha$ \\
\hline
$1\mathbf{0}^p$, $\alpha\mathbf{0}^p$ & $01\mathbf{0}^{\ell+1}$, $0\alpha\mathbf{0}^{\ell+1}$, $1\mathbf{0}^{\ell+2}$, $11\mathbf{0}^{\ell+1}$, $1\alpha\mathbf{0}^{\ell+1}$, $\alpha\mathbf{0}^{\ell+2}$, $\alpha1\mathbf{0}^{\ell+1}$, $\alpha\alpha\mathbf{0}^{\ell+1}$ & $\mathbf{0}^q\alpha^2$ \\
\hline
$\alpha^2\mathbf{0}^p$ & $\mathbf{0}^{\ell+3}$, $00\alpha^2\mathbf{0}^{\ell}$, $0\alpha^2\mathbf{0}^{\ell+1}$, $0\alpha^2\alpha^2\mathbf{0}^{\ell}$, $\alpha^2\mathbf{0}^{\ell+2}$, $\alpha^20\alpha^2\mathbf{0}^{\ell}$, $\alpha^2\alpha^2\mathbf{0}^{\ell+1}$, $\alpha^2\alpha^2\alpha^2\mathbf{0}^{\ell}$ & $\mathbf{0}^q\alpha^2$ \\
\hline
$1\mathbf{0}^p$, $\alpha\mathbf{0}^p$ & $11\mathbf{0}^{\ell+1}$, $11\alpha\mathbf{0}^\ell$, $1\alpha1\mathbf{0}^\ell$, $1\alpha\alpha\mathbf{0}^\ell$, $\alpha11\mathbf{0}^\ell$, $\alpha1\alpha\mathbf{0}^\ell$, $\alpha\alpha1\mathbf{0}^\ell$, $\alpha\alpha\mathbf{0}^{\ell+1}$ & $\mathbf{0}^q1$, $\mathbf{0}^q\alpha$ \\
\hline
\end{tabularx}
\end{table}

\begin{figure}
\centering
\begin{tikzpicture}[
    node distance=3cm,
    circ/.style={circle, draw, minimum size=8mm, inner sep=0pt},
    arr/.style={-Latex, bend left=15},
    looparr/.style={-Latex, out=135, in=45, looseness=8}
]
\node[circ] (A) at (0,0) {};
\node[circ] (B) at (3,0) {};
\node[circ] (C) at (3,3) {};
\node[circ] (D) at (0,3) {};
\draw[looparr, out=315, in=225] (A) to node[above] {$\alpha^2$} (A);
\draw[looparr, out=315, in=225] (B) to node[above] {$\alpha$} (B);
\draw[looparr] (C) to node[below] {$1$} (C);
\draw[looparr] (D) to node[below] {$0$} (D);
\draw[arr] (A) to node[left] {$0$} (D);
\draw[arr] (B) to node[below] {$0$} (D);
\draw[arr] (B) to node[left] {$1$} (C);
\draw[arr] (C) to node[right] {$\alpha$} (B);
\draw[arr] (C) to node[above] {$0$} (D);
\draw[arr] (D) to node[right] {$\alpha^2$} (A);
\draw[arr] (D) to node[below] {$\alpha$} (B);
\draw[arr] (D) to node[above] {$1$} (C);
\end{tikzpicture}
\vspace{-0.3em}
\caption{Finite-state transition diagram for GF$(4)$ RES-LOCO code.}
\label{fig:gf4_state_diagram}
\vspace{-0.5em}
\end{figure}
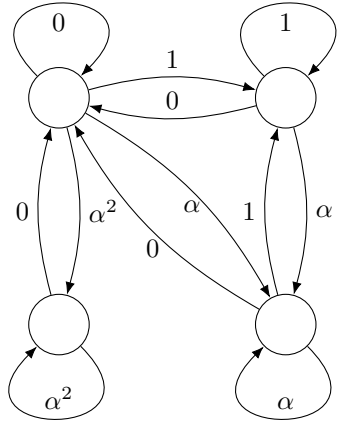

For the case where $\ell = 0$, we calculated finite-length rates for various codeword lengths, using the three bridging schemes described above. These rates are provided in Table~\ref{table_GF_4_capacity}. We show that, in all three cases, the rate approaches the capacity of this constrained coding scheme, which is $0.8323$, as the codeword size grows. Additionally, we observe that the bridging scheme with $\eta_3 = \ell + 2$ consistently performs the best out of the three. Note that the capacity of any constrained code can be systematically obtained from the finite-state transition diagram (FSTD) characterizing it as shown below, and the FSTD of the GF($4$) RES-LOCO code with $\ell=0$ is provided in Fig.~\ref{fig:gf4_state_diagram}.

Since the bridging scheme with $\eta_3 = \ell + 2$ consistently performs the best out of the three. Thus, our rate relation for GF$(4)$ $\ell = 0$ coding is:
\begin{equation}
R_{\text{GF}(4)} = \frac{\lfloor \log_2(N_{4,0}(m)-2) \rfloor + 2}{2 (m+2)}. \label{eq:Rgf4_l0}
\end{equation}
As $m$ goes to $\infty$, we reach the capacity but also the codeword-to-message error propagation factor $\frac{1}{2}(s-1)$ increases, where $\frac{1}{2}(s-1) = \frac{1}{2}(\lfloor \log_2(N_{4,0}(m)-2) \rfloor-1)$. The reason behind this is in Subsection~\ref{subsec_error_prop}. That is why moderate values of $m$ are recommended.

The (normalized) capacity $C$ is equal to $\frac{1}{2}\log_2(\lambda_{\max}) = 0.8323$, where $\lambda_{\max}$ is the maximum real positive eigenvalue of transition matrix $\mathbf{A}$. Note that
\[
\mathbf{A} = \begin{bmatrix}
1 & 1 & 1 & 1 \\
1 & 1 & 1 & 0 \\
1 & 1 & 1 & 0 \\
1 & 0 & 0 & 1
\end{bmatrix},
\]
where rows and columns are associated with $c_{i} = 0, 1, \alpha, \alpha^2$ ordered from top to bottom and from left to right, respectively. Observe that $\mathbf{A}$ is derived from the FSTD in Fig.~\ref{fig:gf4_state_diagram}.

For $m=18$, $R_{\text{GF}(4)}=0.8$, which is already close to the capacity and the error propagation factor is $\frac{1}{2}(\left\lfloor\log_2(N_{4,0}(18) - 2)\right\rfloor-1) = 14.5$, which is acceptable.

\begin{table*}
\caption{Finite-length Rates With the Three Bridging Schemes Versus Codeword Length for \text{GF}($4$) RES-LOCO Codes and $\ell = 0$}
\vspace{-0.5em}
\centering
\scalebox{1.00}
{
\begin{tabular}{|c|c|c|c|c|c|c|c|c|c|c|c|}
\hline \diagbox{$\eta_3$}{$m$} & 6 & 9 & 12 & 15 & 21 & 30 & 45 & 75 & 105 & 170 & 260\\
\hline $1$ & 0.7143 & 0.7500 &	0.7692	& 0.7813  & 0.7955	& 0.8065 &	0.8152	& 0.8224 & 0.8255 & 0.8275 & 0.8295 \\
\hline $2$ & 0.7500	&0.7727&	0.7857	&0.7941&	0.8043&	0.8125	&0.8191&	0.8247&	0.8271&	0.8285&	0.8302\\
\hline $3$ & 0.7222	&0.7500&	0.7667&	0.7778&	0.7917&	0.8030&	0.8125&	0.8205&	0.8241&	0.8266&	0.8289\\
\hline
\end{tabular}}
\label{table_GF_4_capacity}
\vspace{-0.5em}
\end{table*}

\subsection{RES-LOCO \text{GF}($8$) Bridging and Self-Clocking}

\text{GF}($8$) RES-LOCO codes prevent sneak paths of size $1\times 1$, with $\ell = 0$. Similarly to the \text{GF}($4$) case, we consider bridging in three cases: $\eta_3 = 1$, $\eta_3 = 2$, and $\eta_3 = 3$.

In the $\eta_3 = 1$ case, the only option that works for all cases is to bridge with a single $0$, since if the left-hand codeword ends with $\alpha^6$ and the right-hand codeword begins with $\alpha^5$ (or vice versa), inserting any symbol other than $0$ in between the codewords would create a sneak path. Recall that $\alpha^6$ corresponds to $[1 ~ 1 ~ 1 ]^\mathrm{T}$, while $\alpha^5$ corresponds to $[1 ~ 1 ~ 0 ]^\mathrm{T}$. Hence, in this case, we are unable to encode any additional bits within the bridging segment and have to remove the all-zero codeword. Therefore, $\eta_2=0$ and $\eta_1=1$.

In the $\eta_3 = 2$ case, there are at least $4$ compatible bridging patterns (the minimum is $4$)  for every possible ordered pair of codewords to be concatenated, which results in $\eta_2 = 2$. Additionally, we need to remove the all-zero and all-$\alpha^6$ codewords. Therefore, we have $\eta_1 = 2$. The possible bridging arrangements for this case are provided in Table~\ref{table_GF_8_l_2}.
\begin{table}
\caption{Possible Bridging Patterns for \text{GF}($8$) RES-LOCO Codes With $\eta_3 = 2$}
\vspace{-0.5em}
\centering
\scalebox{1.00}
{
\begin{tabular}{|c|c|c|}
\hline Left-hand LOCO& Possible bridging & Right-hand LOCO  \\ codeword ending & scenarios & codeword beginning  \\ with &  & with \\
\hline $\alpha^6$ & $00$, $0\alpha^6$, $\alpha^60$, $\alpha^6\alpha^6$ & $\alpha^6$ \\
\hline $\alpha^6$ & $00$, $01$, $0\alpha^5$, $\alpha^60$ & $\alpha^5$ \\
\hline $\alpha^6$ & $00$, $01$, $0\alpha$, $\alpha^60$ & $\alpha^4$, $\alpha^3$, $\alpha$, $1$, $0$\\
\hline $\alpha^6$ & $00$, $0\alpha^2$, $0\alpha^3$, $\alpha^60$ & $\alpha^2$ \\
\hline $\alpha^5$ & $00$, $0\delta_1$, $10$, $\alpha^50$ & $\delta_1$: $\delta_1 \in \{1,\alpha, \ldots,\alpha^6 \}$ \\
\hline $\alpha^5$ & $00$, $0\alpha^5$, $\alpha^50$, $11$ & $0$ \\
\hline $\alpha^2$ & $00$, $0\delta_2$, $\alpha^20$, $\alpha^30$ & $\delta_2$: $\delta_2 \in \{1,\alpha, \ldots,\alpha^6 \}$ \\
\hline $\alpha^2$ & $00$, $0\alpha^2$, $\alpha^20$, $\alpha^3\alpha^3$ & $0$ \\
\hline $0$, $1$, $\alpha$, $\alpha^3$, $\alpha^4$ & $00$, $10$, $\alpha0$, $\alpha^30$ & Any symbol \\ \hline
\end{tabular}}
\label{table_GF_8_l_2}
\vspace{-0.5em}
\end{table}

In the $\eta_3 = 3$ case, there are at least $8$ compatible bridging patterns (the minimum is $14$)  for every possible ordered pair of codewords to be concatenated, which results in $\eta_2 = 3$. For some cases, the number of bridging scenarios is limited to a number greater than or equal to $8$ but strictly less than $16$, so $\eta_2$ cannot be larger than $3$. Hence, an effective bridging scheme would be to simply use the arrangement $0\delta_30$, where $\delta_3$ can be any of the \text{GF}($8$) symbols. We also need to remove the all-zero codeword. Therefore, we have $\eta_1 = 1$.

Similarly to our \text{GF}($4$) code, we calculated finite-length rates for various \text{GF}($8$) RES-LOCO codeword lengths using the three bridging schemes described above. These rates are provided in Table~\ref{table_GF_8_capacity}. In all three cases, the rate approaches the capacity of this constrained coding scheme, which is $0.8145$, as the codeword size grows. Additionally, we observe that the bridging scheme with $\eta_3 = 1$ consistently performs the best out of the three, even though no new bits of information are coded within the bridging segment.

As we discussed before, the best bridging scheme rate-wise is adding $0$ between two codewords. We remove all $0$'s codeword due to level-based signaling. Thus, our rate relation for GF$(8)$ coding is:
\begin{equation}
R_{\text{GF}(8)} = \frac{\lfloor \log_2(N_{8,0}(m)-1) \rfloor}{3 (m+1)}. \label{eq:25}
\end{equation}
As $m$ goes to $\infty$, we reach the capacity but also the codeword-to-message error propagation factor $\frac{1}{2}(s-1)$ increases, where $\frac{1}{2}(s-1) = \frac{1}{2}(\lfloor \log_2(N_{8,0}(m)-1) \rfloor-1)$. That is why moderate values of $m$ are recommended.

\begin{figure}
\centering
\begin{tikzpicture}[
    scale=0.83, transform shape,
    >=latex,
    auto,
    every edge/.style={
        draw, ->, thick
    }
]
    \node[circle, draw=blue!70, fill=blue!25, thick, minimum size=1cm, font=\sffamily\large] (S1) at ({2.5*cos(90 - (1-1)*45)}, {4*sin(90 - (1-1)*45)}) {$0$};
    \node[circle, draw=blue!70, fill=blue!25, thick, minimum size=1cm, font=\sffamily\large] (S2) at ({2.5*cos(90 - (2-1)*45)}, {4*sin(90 - (2-1)*45)}) {$1$};
    \node[circle, draw=blue!70, fill=blue!25, thick, minimum size=1cm, font=\sffamily\large] (S3) at ({2.5*cos(90 - (3-1)*45)}, {4*sin(90 - (3-1)*45)}) {$\alpha$};
    \node[circle, draw=blue!70, fill=blue!25, thick, minimum size=1cm, font=\sffamily\large] (S4) at ({2.5*cos(90 - (4-1)*45)}, {4*sin(90 - (4-1)*45)}) {$\alpha^2$};
    \node[circle, draw=blue!70, fill=blue!25, thick, minimum size=1cm, font=\sffamily\large] (S5) at ({2.5*cos(90 - (5-1)*45)}, {4*sin(90 - (5-1)*45)}) {$\alpha^3$};
    \node[circle, draw=blue!70, fill=blue!25, thick, minimum size=1cm, font=\sffamily\large] (S6) at ({2.5*cos(90 - (6-1)*45)}, {4*sin(90 - (6-1)*45)}) {$\alpha^4$};
    \node[circle, draw=blue!70, fill=blue!25, thick, minimum size=1cm, font=\sffamily\large] (S7) at ({2.5*cos(90 - (7-1)*45)}, {4*sin(90 - (7-1)*45)}) {$\alpha^5$};
    \node[circle, draw=blue!70, fill=blue!25, thick, minimum size=1cm, font=\sffamily\large] (S8) at ({2.5*cos(90 - (8-1)*45)}, {4*sin(90 - (8-1)*45)}) {$\alpha^6$};
    
    \scope[every path/.style={->, thick, loop, distance=1.2cm}]
        \path (S1) edge [out=120, in=60] (S1);
        \path (S2) edge [out=75, in=15] (S2);
        \path (S3) edge [out=45, in=-45] (S3);
        \path (S4) edge [out=-15, in=-75] (S4);
        \path (S5) edge [out=-60, in=-120] (S5);
        \path (S6) edge [out=-105, in=-165] (S6);
        \path (S7) edge [out=225, in=135] (S7);
        \path (S8) edge [out=165, in=105] (S8);
    \endscope
    
    \scope[every path/.style={->, thick}]
        \path (S1) edge [bend right=10] (S2); 
        \path (S1) edge [bend right=20] (S3); 
        \path (S1) edge [bend right=20] (S4); 
        \path (S1) edge [bend right=20] (S5);
        \path (S1) edge [bend left=20] (S6); 
        \path (S1) edge [bend left=20] (S7); 
        \path (S1) edge [bend left=10] (S8);
        \path (S2) edge [bend right=10] (S1); 
        \path (S2) edge [bend right=10] (S3); 
        \path (S2) edge (S5); 
        \path (S2) edge [bend left=10] (S6); 
        \path (S2) edge [bend left=20] (S7);
        \path (S3) edge [bend right=15] (S1); 
        \path (S3) edge [bend right=10] (S2); 
        \path (S3) edge [bend left=10] (S5); 
        \path (S3) edge [bend left=15] (S6);
        \path (S4) edge [bend right=15] (S1); 
        \path (S4) edge (S5);
        \path (S5) edge [bend right=20] (S1); 
        \path (S5) edge (S2); 
        \path (S5) edge [bend left=10] (S3); 
        \path (S5) edge (S4); 
        \path (S5) edge [bend left=10] (S6);
        \path (S6) edge [bend left=20] (S1); 
        \path (S6) edge [bend left=10] (S2); 
        \path (S6) edge [bend right=10] (S3); 
        \path (S6) edge [bend right=10] (S5);
        \path (S7) edge [bend left=15] (S1); 
        \path (S7) edge [bend left=10] (S2);
        \path (S8) edge [bend left=10] (S1);
    \endscope
\end{tikzpicture}
\vspace{-1.5em}
\caption{Finite-state transition diagram for GF$(8)$ RES-LOCO code.}
\label{fig:gf8_state_diagram}
\end{figure}
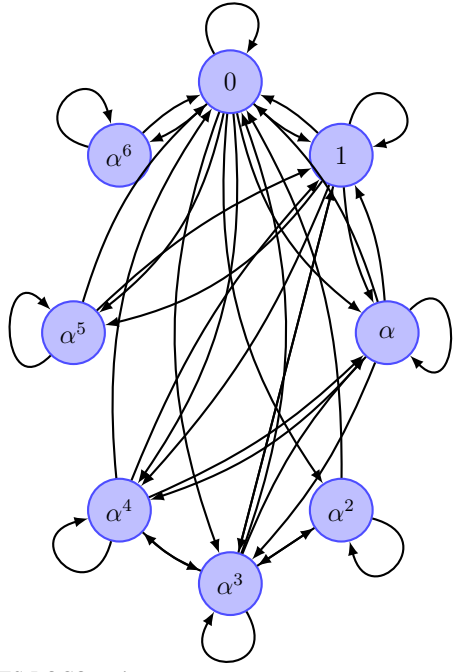

The (normalized) capacity $C$ is equal to $\frac{1}{3}\log_2(\lambda_{\max}) = 0.8145$, where $\lambda_{\max}$ is the maximum real positive eigenvalue of transition matrix $\mathbf{A}$. Note that
\[
\mathbf{A} = \begin{bmatrix}
1 & 1 & 1 & 1 & 1 & 1 & 1 & 1 \\
1 & 1 & 1 & 0 & 1 & 1 & 1 & 0 \\
1 & 1 & 1 & 0 & 1 & 1 & 0 & 0 \\
1 & 0 & 0 & 1 & 1 & 0 & 0 & 0 \\
1 & 1 & 1 & 1 & 1 & 1 & 0 & 0 \\
1 & 1 & 1 & 0 & 1 & 1 & 0 & 0 \\
1 & 1 & 0 & 0 & 0 & 0 & 1 & 0 \\
1 & 0 & 0 & 0 & 0 & 0 & 0 & 1
\end{bmatrix},
\]
where rows and columns are associated with $c_{i} = 0, 1, \alpha, \ldots, \alpha^6$ ordered from top to bottom and from left to right, respectively. Observe that $\mathbf{A}$ is derived from the FSTD in Fig.~\ref{fig:gf8_state_diagram}.

For $m=18$, $R_{\text{GF}(8)}=0.7719$, which is already close to the capacity and the error propagation factor in this case is $\frac{1}{2}(\left\lfloor\log_2(N_{8,0}(18) - 1)\right\rfloor-1) = 21.5$, which is acceptable.

\begin{table*}
\caption{Finite-length Rates With the Three Bridging Schemes Versus Codeword Length for \text{GF}($8$) RES-LOCO Codes and $\ell = 0$}
\vspace{-0.5em}
\centering
\scalebox{1.00}
{
\begin{tabular}{|c|c|c|c|c|c|c|c|c|c|c|c|}
\hline \diagbox{$\eta_3$}{$m$} & 6 & 9 & 13 & 18 & 22 & 31 & 40 & 70 & 110 & 160 & 250\\
\hline $1$ & 0.7143&	0.7333&	0.7619&	0.7719&	0.7826&	0.7917&	0.7967&	0.8028&	0.8078&	0.8095&	0.8114 \\
\hline $2$ & 0.7083	&0.7273&	0.7556&	0.7667&	0.7778&	0.7879&	0.7937	&0.8009&	0.8065&	0.8086&	0.8108\\
\hline $3$ & 0.6667&	0.6944&	0.7292&	0.7460&	0.7600&	0.7745&	0.7829&	0.7945&	0.8024&	0.8057&	0.8090\\
\hline
\end{tabular}}
\label{table_GF_8_capacity}
\vspace{-0.5em}
\end{table*}

\section{Run-Length Limited Code Design for Sneak-Path Reduction}\label{sec_rll}

For the $b=4$ case, a non-binary LOCO code designed to mitigate sneak-path effects will be defined over GF($16$). The same applies also to the case of $b > 4$. In this case, the complexity remarkably increases for multiple reasons, including the notable increase in the number of patterns to forbid, the higher storage overhead because of bigger cardinalities, and the more sophisticated encoding-decoding rule. Instead, we have turned to run-length-limited (RLL) coding solutions, which while not being rate-wise optimal, they still offer relatively high coding rate as well as low complexity. We have designed a binary RLL code, where we code vertically on the crossbar array, for the $b = 4$ case. We also modify a prior result from literature to combat the case with a greater number of consecutive rows read simultaneously.

\subsection{Vertical RLL Coding for Groups of Four Consecutive Rows}

In the case where four consecutive rows of memristors are grouped and read at once with all other rows grounded, we have designed a binary RLL coding scheme to reduce the number of sneak paths. In this scheme, the data over every $4$-element column is treated as an RLL codeword, and consecutive $1$'s are prevented as the forbidden pattern becomes $11$. This results in a code with a rate of $0.75$ that prevents all sneak paths of circumference $4$ within the $4$-row segment, since every possible arrangement of a circumference-$4$ sneak path would have to feature two vertical consecutive $1$'s. Out of $16$ possible $4$-bit vertical codewords, $8$ are free of the forbidden pattern $11$, allowing $\log_2 8 = 3$ bits to be encoded per $4$-bit codeword, and thus resulting in a rate of $0.75$. These codewords are the following: $[0 ~ 0 ~ 0 ~ 0]^{\mathrm{T}}$, $[0 ~ 0 ~ 0 ~ 1]^{\mathrm{T}}$, $[0 ~ 0 ~ 1 ~ 0]^{\mathrm{T}}$, $[0 ~ 1 ~ 0 ~ 0]^{\mathrm{T}}$, $[0 ~ 1 ~ 0 ~ 1]^{\mathrm{T}}$, $[1 ~ 0 ~ 0 ~ 0]^{\mathrm{T}}$, $[1 ~ 0 ~ 0 ~ 1]^{\mathrm{T}}$, $[1 ~ 0 ~ 1 ~ 0]^{\mathrm{T}}$. Observe that preventing $11$ vertically (or horizontally) eliminates all sneak paths of circumference $4$. The main advantage this scheme provides is its low decoding complexity, as the decoding is performed using a small lookup table with only $8$ entries. On the other hand, its main disadvantage is its rate of $0.75$, which is lower than those of our RES-LOCO codes.

\subsection{General RLL Solution}

In \cite{nguyen2023locally}, a binary RLL coding solution is presented, which prevents sneak paths from occurring in every sliding window of size $\mu_1\times \mu_1$ in an $\mu_2\times \mu_2$ memristor crossbar array, using $(\mu_1-1,\infty)$ RLL codewords over the rows of the two dimensional array. Such a code imposes at least ($\mu_1-1$) $0$'s between consecutive $1$'s. For $\mu_1 > 1$, this coding scheme has rates that are quite low for storage. For example, the capacity drops below $0.5$ as $\mu_1$ becomes $3$ or higher. We can modify this scheme to focus on the most detrimental sneak paths, those of circumference $4$, as follows. Set the data over each row of the 2D resistive array to store a $(1,\infty)$ RLL codeword, preventing the pattern $11$ from occurring anywhere within a codeword. This code prevents all sneak paths of circumference $4$ over the whole array, but has the lowest rate of all the codes we have discussed so far, as its capacity is $0.6942$. Having said that, despite the lowered rate, this code provides an effective solution to prevent sneak paths of circumference $4$ in crossbar arrays where at least five rows are read simultaneously while the others are grounded, situations where enumerative constrained coding solutions can grow notably in complexity.

\section{Simulation Results}\label{sec_sims}

In this section, we present simulation results that demonstrate the effectiveness of our coding schemes in mitigating the sneak-path problem.

\subsection{GF$(4)$ Coding Results}

Simulations were conducted for GF$(4)$ RES-LOCO codes with different $\ell$ values ($0$, $1$, and $2$), as these correspond to the most dominant sneak-path errors. For larger $\ell$ values, line (wire) resistance becomes more dominant compared with smaller $\ell$ values, reducing the probability of sneak-path occurrence as circumference increases. We also show how using RES-LOCO codes with specific $\ell$ values contributes to the mitigation of sneak-paths of circumference more than $2\ell+4$.

\subsubsection{Simulation Setup~1: All-Zero Row Insertion}
In a memristor array with two rows, sneak paths with circumference $4$ are not observed when the data is coded using a GF$(4)$ RES-LOCO code. However, for arrays with more than two rows, sneak paths with circumference $4$ and higher begin to appear even when the data is coded. To prevent this, an all-zero row is inserted after every second data row (i.e., as the $3$rd, $6$th, $9$th, etc., row). This strategy protects the data stored in each group of two rows by blocking the most dominant type of sneak path, which has a circumference of $4$.

Let $S_L$ denote the number of sneak paths with circumference $L$ averaged over all Monte Carlo trials. We performed the simulation using a message length of $18$ and $10^6$ Monte Carlo trials. The number of rows was fixed at $98$, while the number of columns was set to $98$, $102$, and $106$ for $\ell$ values of $0$, $1$, and $2$, respectively. Recall that for $\ell=0,1,2$, we have $L_{\max}=4,6,8$, respectively. The simulation results are presented in Table~\ref{tab:Sim1Gf4}.

\begin{table}[h]
\centering
\caption{Simulation Setup~1 Results: All-Zero Row Insertion}
\begin{tabular}{|l|c|c|c|c|}
\hline
\diagbox{$L_{\max}$}{$S_k$} & $S_4$ & $S_6$ & $S_8$ & $S_{10}$ \\
\hline
$4$ & $0$ & $511.70$ & $1226.7$ & $1599.8$ \\
\hline
$6$ & $0$ & $284.54$ & $880.74$ & $1142.6$ \\
\hline
$8$ & $0$ & $279.98$ & $824.46$ & $1065.7$ \\
\hline
\end{tabular}
\label{tab:Sim1Gf4}
\end{table}

Although achieving $S_4=0$ is a positive result, the overall code rate decreases due to the insertion of redundant all-zero rows as one all-zero row is added for every two data rows. Consequently, the new effective rate, $R_{\text{new}}$, is two-thirds of the original rate, $R$:
\begin{equation}
R_{\text{new}} = \frac{2}{3}R.
\end{equation}
For example, with the original rate $R = 0.7143$ (for $L_{\max}=6$ and message length $18$), the new rate becomes:
\begin{equation}
R_{\text{new}} = \frac{2}{3} \times 0.7143 = 0.4762.
\end{equation}

We note that these reported numbers for $S_4$, $S_6$, $S_8$, and $S_{10}$, generated while using our GF$(4)$ RES-LOCO codes at various $L_{\max}$ values, are remarkably lower than those associated with the uncoded setting even when the all-zero rows are used.

\subsubsection{Simulation Setup~2: Sequential Paired Reading}
In Simulation Setup~1, achieving $S_4=0$ comes at the cost of reducing the code rate to $2/3$ of its original value due to the insertion of redundant, all-zero rows. To address this rate loss, we adopt a method from \cite{Cassuto2016Information} that eliminates the need for all-zero rows by modifying the data reading process. In this method, the memristor array is read in non-overlapping pairs. Specifically, Rows $(2k-1)$ and $2k$ are read together at each step, where $k \in \{1, 2, \ldots\}$. For instance, Rows $1$ and $2$ are read first, followed by Rows $3$ and $4$, and so on. When using our GF$(4)$ RES-LOCO codes, this technique successfully prevents the dominant $S_4$ sneak paths while preserving the original code rate. The same is also true for all $S_L$ sneak paths when the RES-LOCO code is adjusted accordingly via the correct $L_{\max}$.

The simulation was conducted with a message length of $18$ and $10^6$ Monte Carlo trials. The number of rows was fixed at $98$, while the number of columns was set to $98$, $102$, and $106$ for $L_{\max}=4$, $L_{\max}=6$, and $L_{\max}=8$, respectively.

The results are presented in Table~\ref{tab:Sim2Gf4}. Observe that Uncoded~1 is associated with array dimensions $98 \times 98$, Uncoded~2 is associated with array dimensions $98 \times 102$, and Uncoded~3 is associated with the array dimensions $98 \times 106$. Hence, Uncoded~1 setting is compared with GF($4$) RES-LOCO coding where $\ell=0$, Uncoded~2 setting is compared with GF($4$) RES-LOCO coding where $\ell=1$, and Uncoded~3 setting is compared with GF($4$) RES-LOCO coding where $\ell=2$. Observe also that for all uncoded settings, we use $p_1=0.5$ for unconstrained user messages.

\begin{table}[h]
\centering
\caption{Simulation Setup~2 Results: Sequential Paired Reading}
\begin{tabular}{|l|c|c|c|c|}
\hline
\diagbox{$L_{\max}$}{$S_k$} & $S_4$ & $S_6$ & $S_8$ & $S_{10}$ \\
\hline
Uncoded~1 & $2352.3$ & $4656$ & $6911.5$ & $9119$ \\
\hline
$4$ & $0$ & $281.17$ & $358.72$ & $391.75$ \\
\hline
Uncoded~2 & $2449.3$ & $4849$ & $7199.5$ & $9501$ \\
\hline
$6$ & $0$ & $0$ & $71.93$ & $100.04$ \\
\hline
Uncoded~3 & $2546.3$ & $5042$ & $7487.5$ & $9883$ \\
\hline
$8$ & $0$ & $0$ & $0$ & $19.82$ \\
\hline
\end{tabular}
\label{tab:Sim2Gf4}
\end{table}

The results demonstrate clear improvement over Simulation Setup~1, both in terms of sneak-path reduction and code rate preservation. However, this approach introduces a potential trade-off in the form of increased latency. Because the array must be read sequentially (pair by pair), the total reading time increases compared with reading the entire array simultaneously. Furthermore, power consumption in this method can potentially be higher.

A detailed quantitative analysis of the latency issue is beyond the scope of this work. Nevertheless, we suggest that for small-scale memristor arrays, this effect would be manageable. Furthermore, even for large-scale arrays, the impact may not be critical, depending on the application requirements and tolerance for processing delays.

We note that these reported numbers for $S_4$, $S_6$, $S_8$, and $S_{10}$, generated while using our GF$(4)$ RES-LOCO codes at various $L_{\max}$ values, are remarkably lower than those associated with the uncoded setting. In particular, the reduction factors are $\infty$, $16.56$, $19.26$, and $23.27$ for $S_4$, $S_6$, $S_8$, and $S_{10}$, respectively, when the array dimensions are $98 \times 98$. The reduction factors are $\infty$, $\infty$, $100.09$, and $94.97$ for $S_4$, $S_6$, $S_8$, and $S_{10}$, respectively, when the array dimensions are $98 \times 102$. The reduction factors are $\infty$, $\infty$, $\infty$, and $498.64$ for $S_4$, $S_6$, $S_8$, and $S_{10}$, respectively, when the array dimensions are $98 \times 106$.

Observe that for the case of the $98 \times 106$ crossbar array, where our GF$(4)$ RES-LOCO code has $\ell=2$, to achieve the coded averages of $S_4$, $S_6$, and $S_8$ in the uncoded setting, the value of $p_1$ must be $1$ or $0$, which implies that no information can be stored. As for $S_{10}$, to achieve the coded average in the uncoded setting, the value of $p_1$ must be $0.99987$ or $0.05093$, which implies that limited amount of information can be stored. Furthermore, to achieve the cumulative coded average of $S_4$, $S_6$, $S_8$, and $S_{10}$ in the uncoded setting, the value of $p_1$ must be $0.99995$ or $0.03722$, which implies even further limitations on the amount of information to be stored. These are obtained via Lemma~\ref{lemma:exp_sp_l} and Lemma~\ref{lemma:exp_sp_l_max} relations.

\subsubsection{Simulation Setup~3: Simultaneous Reading}
Simulation Setup~3 involves simultaneous reading, where all rows of the memristor crossbar array are read at once (with no all-zero rows and no grounding).

This method has two key characteristics:
\begin{enumerate}
    \item Reduced latency: By reading all rows simultaneously, this approach avoids the sequential reading latency of Simulation Setup~2.
    \item Preserved rate: The code rate is identical to that of Simulation Setup~2, as no redundant rows are used.
\end{enumerate}

The simulation parameters are identical to those of Simulation Setup~2. The results are presented in Table~\ref{tab:Sim3Gf4}.

\begin{table}[h]
\centering
\caption{Simulation Setup~3 Results: Simultaneous Reading}
\begin{tabular}{|l|c|c|c|c|}
\hline
\diagbox{$L_{\max}$}{$S_k$} & $S_4$ & $S_6$ & $S_8$ & $S_{10}$ \\
\hline
$4$ & $484.78$ & $1762.0$ & $2843.6$ & $3855.8$ \\
\hline
$6$ & $427.61$ & $1252.2$ & $2109.3$ & $2922.8$ \\
\hline
$8$ & $420.21$ & $1236.6$ & $1989.7$ & $2728.6$ \\
\hline
\end{tabular}
\label{tab:Sim3Gf4}
\end{table}

While this approach avoids both additional latency and rate reduction, it has a significant drawback: the number of sneak paths with circumference $4$ ($S_4$) cannot be zero using our coding techniques. Observe that sneak paths of circumference $4$ can be entirely removed under simultaneous reading using the RLL coding idea if we bridge with a $0$ vertically after each binary $4$-tuple. However, this approach is conceptually similar to Simulation Setup~1. 

To properly evaluate the effectiveness of our proposed method (Simulation Setup~3), we perform comparisons in the performance analysis part below, comparing observed cumulative sneak-path average values in simulations against theoretically expected values calculated under maxentropic probability assumptions. Maxentropic probabilities are those associated with maximum entropy of the FSTD, and they are obtained as discussed in \cite{marcus_book} and \cite{ahh_simpleloco}.

\subsubsection{Performance Analysis: Comparison with Theoretical Expectation}

To evaluate the performance of our method, we conduct a stringent comparison, one that would demonstrate the minimum gains that can be achieved using our coding schemes. In particular, we measure the total number of observed sneak paths from the simultaneous reading simulations (Simulation Setup~3) averaged over all trials, $\sum_k S_k$, and compare it against the theoretically expected value of $\sum_k S_k$, which we denote by $\mathbb{E}[\sum_k S_k]$, obtained via Lemma~\ref{lemma:exp_sp_l_max}.

The ``Total observed SPs'' column in Table~\ref{tab:s4_comparison} represents the cumulative sum of sneak paths with circumferences up to the given $L_{\max}$. Specifically,
\begin{itemize}
    \item For $L_{\max}=4$: It is $S_4$.
    \item For $L_{\max}=6$: It is $S_4 + S_6$.
    \item For $L_{\max}=8$: It is $S_4 + S_6 + S_8$.
\end{itemize}

The improvement, or the reduction, factor is calculated by dividing the theoretical expectation of $\sum_k S_k$ by this cumulative observed count. The complete comparison is summarized in Table~\ref{tab:s4_comparison}.

\begin{table}[h]
\centering
\caption{Comparison of Total Observed SPs With Theoretical $\sum_k S_k$ Expectation}
\label{tab:s4_comparison}
\small
\setlength{\tabcolsep}{4pt}
\renewcommand{\arraystretch}{1.2}
\begin{tabular}{|c|cc|c|c|c|}
\hline
\textbf{$L_{\max}$} & \textbf{$p_0$} & \textbf{$p_1$} & \thead{$\mathbb{E}[\sum_k S_k]$ \\ (theory)} & \thead{Total observed SPs \\ $\sum S_k$} & \thead{Impr. \\ Factor} \\
\hline
$4$ & $0.6528$ & $0.3472$ & $1028.3$ & $484.78$ & $2.121$ \\
\hline
$6$ & $0.6680$ & $0.3320$ & $2854.5$ & $1679.81$ & $1.699$ \\
\hline
$8$ & $0.6674$ & $0.3326$ & $5923.1$ & $3646.51$ & $1.624$ \\
\hline
\end{tabular}
\end{table}

As shown in Table~\ref{tab:s4_comparison}, even using this conservative metric, the improvement factor exceeds $1.6$ for all cases. This strongly suggests that with other simulation setups that involve circuit plus coding solutions, the improvement factor will be remarkably higher as shown above. Observe that these calculations do not account for wire resistance. Including wire resistance in the model would likely decrease the number of observed sneak paths, particularly those with large circumference.

\subsection{GF$(8)$ Coding Results}

Simulations were conducted for GF$(8)$ RES-LOCO codes with $\ell = 0$ (i.e., $L_{\max} = 4$), as this corresponds to the most dominant sneak-path errors. 

\subsubsection{Simulation Setup~1: All-Zero Row Insertion}
In a memristor array with three rows, sneak paths with circumference $4$ are not observed when the data is coded using a GF$(8)$ RES-LOCO code. However, for arrays with more than three rows, sneak paths with circumference $4$ and higher begin to appear even when the data is coded. To prevent this, an all-zero row is inserted after every third data row (i.e., as the $4$th, $8$th, $12$th, etc., row). This strategy protects the data stored in each group of three rows by blocking the most dominant type of sneak path, which has a circumference of $4$.

Recall that $S_L$ denotes the number of sneak paths with circumference $L$ averaged over all Monte Carlo trials. We performed the simulation using a message length of $18$ and $10^6$ Monte Carlo trials. The number of rows was fixed at $99$, while the number of columns was set to $94$ for $\ell=0$. Recall that for $\ell=0$, we have $L_{\max}=4$. The simulation results are:
\begin{equation}
S_4 = 0, \quad S_6 = 760.07, \quad S_8 = 1175.7, \quad S_{10} = 1865.9.
\end{equation}

Although achieving $S_4=0$ is a positive result, the overall code rate decreases due to the insertion of redundant all-zero rows as one all-zero row is added for every three data rows. Consequently, the new effective rate, $R_{\text{new}}$, is three-quarters of the original rate, $R$:
\begin{equation}
R_{\text{new}} = \frac{3}{4}R.
\end{equation}
For example, with the original rate $R=0.7719$ (for $L_{\max}=4$ and message length $18$), the new rate becomes:
\begin{equation}
R_{\text{new}} = \frac{3}{4} \times 0.7719 = 0.5789.
\end{equation}

We note that these reported numbers for $S_4$, $S_6$, $S_8$, and $S_{10}$, generated while using our GF$(8)$ RES-LOCO codes at $L_{\max}=4$, are remarkably lower than those associated with the uncoded setting even when the all-zero rows are used.

\subsubsection{Simulation Setup~2: Sequential Triplet Reading}
In Simulation Setup~1, achieving $S_4=0$ comes at the cost of reducing the code rate to $3/4$ of its original value due to the insertion of redundant, all-zero rows. To address this rate loss, we adopt a method from \cite{Cassuto2016Information} that eliminates the need for all-zero rows by modifying the data reading process. In this method, the memristor array is read in non-overlapping triplets. Specifically, Rows $(3k-2)$ through $3k$ are read together at each step, where $k \in \{1, 2, \ldots\}$. For instance, Rows $1$ through $3$ are read first, followed by Rows $4$ through $6$, and so on. When using our GF$(8)$ RES-LOCO codes, this technique successfully prevents the dominant $S_4$ sneak paths while preserving the original code rate. 

The simulation was conducted with a message length of $18$ and $10^6$ Monte Carlo trials. The number of rows was fixed at $99$, while the number of columns was set to $94$ for $L_{\max}=4$.

The results are presented in Table~\ref{tab:Sim2Gf8}. Observe that Uncoded~1 is associated with array dimensions $99 \times 94$. Hence, Uncoded~1 setting is compared with GF($8$) RES-LOCO coding where $\ell=0$. Observe also that for the uncoded setting, we use $p_1=0.5$ for unconstrained user messages.

\begin{table}[h]
\centering
\caption{Simulation Setup~2 Results: Sequential Triplet Reading for GF($8$)}
\begin{tabular}{|l|c|c|c|c|}
\hline
\diagbox{$L_{\max}$}{$S_k$} & $S_4$ & $S_6$ & $S_8$ & $S_{10}$ \\
\hline
Uncoded~1 & $2278.5$ & $4509.3$ & $6692.5$ & $8828.5$ \\
\hline
$4$ & $0$ & $638.31$ & $735.87$ & $753.03$ \\
\hline
\end{tabular}
\label{tab:Sim2Gf8}
\end{table}

The results demonstrate clear improvement over Simulation Setup~1, both in terms of sneak-path reduction and code rate preservation. The reason behind the performance gains is that for Simulation Setup~1, we can have sneak paths of circumference more than $4$ showing across different groups of $3$ rows, while this is not possible in Simulation Setup~2 because of the grounding mechanism. However, this approach introduces a potential trade-off in the form of increased latency. Because the array must be read sequentially (triplet by triplet), the total reading time increases compared with reading the entire array simultaneously. Furthermore, power consumption in this method can potentially be higher.

A detailed quantitative analysis of the latency issue is beyond the scope of this work. Nevertheless, we suggest that for small-scale memristor arrays, this effect would be manageable. Furthermore, even for large-scale arrays, the impact may not be critical, depending on the application requirements and tolerance for processing delays.

We note that these reported numbers for $S_4$, $S_6$, $S_8$, and $S_{10}$, generated while using our GF$(8)$ RES-LOCO codes at $L_{\max}=4$, are remarkably lower than those associated with the uncoded setting. In particular, the reduction factors are $\infty$, $7.06$, $9.09$, and $11.72$ for $S_4$, $S_6$, $S_8$, and $S_{10}$, respectively, when the array dimensions are $99 \times 94$.

Observe that for the case of the $99 \times 94$ crossbar array, where our GF$(8)$ RES-LOCO code has $\ell=0$, to achieve the coded average of $S_4$ in the uncoded setting, the value of $p_1$ must be $1$ or $0$, which implies that no information can be stored. Furthermore, to achieve the cumulative coded average of $S_4$, $S_6$, $S_8$, and $S_{10}$ in the uncoded setting, the value of $p_1$ must be $0.9939$ or $0.1949$, which implies that limited amount of information can be stored.

\subsubsection{Simulation Setup~3: Simultaneous Reading}
Simulation Setup~3 involves simultaneous reading, where all rows of the memristor crossbar array are read at once (with no all-zero rows and no grounding).

This method has two key characteristics:
\begin{enumerate}
    \item Reduced latency: By reading all rows simultaneously, this approach avoids the sequential reading latency of Simulation Setup~2.
    \item Preserved rate: The code rate is identical to that of Simulation Setup~2, as no redundant rows are used (preserving the original code rate of $0.7719$).
\end{enumerate}

The simulation parameters are identical to those of Simulation Setup~2. The results are:
\begin{equation}
S_4 = 368.67, \quad S_6 = 1462.5, \quad S_8 = 2462.8, \quad S_{10} = 3335.8.
\end{equation}

While this approach avoids both additional latency and rate reduction, it has a significant drawback: the number of sneak paths with circumference $4$ ($S_4$) cannot be zero using our coding techniques. Observe that sneak paths of circumference $4$ can be entirely removed under simultaneous reading using the RLL coding idea if we bridge with a $0$ vertically after each binary $4$-tuple. However, this approach is conceptually similar to Simulation Setup~1. 

To properly evaluate the effectiveness of our proposed method (Simulation Setup~3), we perform comparisons in the performance analysis part below, comparing the observed average number of sneak paths with circumference $4$ ($S_4$) in simulations against theoretically expected values calculated under maxentropic probability assumptions. Maxentropic probabilities are those associated with maximum entropy of the FSTD, and they are obtained as discussed in \cite{marcus_book} and \cite{ahh_simpleloco}.

\subsubsection{Performance Analysis: Comparison with Theoretical Expectation}

To evaluate the performance of our method, we conduct a stringent comparison, one that would demonstrate the minimum gains that can be achieved using our coding schemes. In particular, we measure the observed number of sneak paths with circumference $4$ from the simultaneous reading simulations (Simulation Setup~3) averaged over all trials, $S_4$, and compare it against its theoretically expected value, which we denote by $\mathbb{E}[S_4]$, obtained via Lemma~\ref{lemma:exp_sp_l_max}.

First, we calculate the maxentropic probabilities of the low-resistance state ($p_1$) and high-resistance state ($p_0$) via the code's transition structure. For the GF$(8)$ code with $L_{\max} = 4$, we find $p_0 = 0.6439$ and $p_1 = 0.3561$. These probabilities are used to calculate the theoretical expectation for an array with $99$ rows and $94$ columns.

The improvement, or the reduction, factor is calculated by dividing the theoretical expectation of $S_4$ by the observed simulation count. The complete comparison is summarized in Table~\ref{tab:s4_comparison_gf8}.

\begin{table}[h]
\centering
\caption{Comparison of Observed $S_4$ With Theoretical Expectation for GF($8$)}
\label{tab:s4_comparison_gf8}
\small
\setlength{\tabcolsep}{4pt}
\renewcommand{\arraystretch}{1.2}
\begin{tabular}{|c|cc|c|c|c|}
\hline
\textbf{$L_{\max}$} & \textbf{$p_0$} & \textbf{$p_1$} & \thead{$\mathbb{E}[S_4]$ \\ (theory)} & \thead{Observed $S_4$ \\ (simulation)} & \thead{Impr. \\ Factor} \\
\hline
$4$ & $0.6439$ & $0.3561$ & $1060$ & $368.67$ & $2.875$ \\
\hline
\end{tabular}
\end{table}

As shown in Table~\ref{tab:s4_comparison_gf8}, even using this conservative metric, the improvement factor is approximately $2.9$. This strongly suggests that with other simulation setups that involve circuit plus coding solutions, the improvement factor will be remarkably higher as shown above. Observe that these calculations do not account for wire resistance. Including wire resistance in the model would likely decrease the number of observed sneak paths, particularly those with large circumference.

\subsection{Error Propagation Analysis}\label{subsec_error_prop}

LOCO codes do not suffer from codeword-to-codeword error propagation. However, codeword-to-message error propagation is an issue. That is, one error in the codeword can result in multiple errors in the message after constrained decoding \cite{Hareedy2022LOCO}. The extent of error propagation is a critical factor in determining a suitable message length. A fundamental trade-off exists in this selection:
\begin{itemize}
    \item Short message lengths limit the propagation of errors, but their corresponding code rates are typically far from the capacity.
    \item Long message lengths can achieve higher rates closer to capacity, but they are more susceptible to higher error propagation.
\end{itemize}

The best case scenario is that one codeword error results in one message error, and the worst case scenario is that one codeword error results in $s$ message errors. In the first, there is no error propagation, while in the second, a single error propagates to $(s-1)$ message bits. Therefore, a quantitative measure for the average extent of error propagation is given by:
\begin{equation} \label{eq:error_prop}
    \text{Error Propagation} = \frac{1}{2}(s-1) = \frac{1}{2} (\left\lfloor \log_2(N(m) - \eta_1) \right\rfloor-1),
\end{equation}
where $N(m)$ is the cardinality of the code with message length $m$, while $s$ is the message length in bits. This metric quantifies the average number of bits that can be affected when a single symbol error occurs during decoding. Table~\ref{tab:error} shows the error propagation values for different code configurations and message lengths.

Note that this trade-off is not relevant for metrics like frame error rate (FER) as it can only be affected by codeword-to-codeword error propagation, and LOCO codes do not suffer from this problem, unlike codes based on finite-state machines. Table~\ref{tab:error} shows that the error propagation factors of RES-LOCO codes are acceptable.

\begin{table}[h]
\centering
\caption{Error Propagation for Different Code Configurations}
\begin{tabular}{|l|c|c|c|c|}
\hline
\diagbox{Configuration}{$m$} & $5$ & $10$ & $15$ & $20$ \\
\hline
GF$(4)$, $\ell=0$ & $3.5$ & $7.5$ & $12$ & $16$ \\
\hline
GF$(4)$, $\ell=1$ & $3.5$ & $7.5$ & $11.5$ & $15.5$ \\
\hline
GF$(4)$, $\ell=2$ & $3.5$ & $7$ & $11$ & $15$ \\
\hline
GF$(8)$, $\ell=0$ & $5.5$ & $11.5$ & $18$ & $24$ \\
\hline
\end{tabular}
\label{tab:error}
\end{table}

\section{Conclusions and Future Work}\label{sec_conc}

We provided estimates for the number of sneak paths in the emerging in-memory computing systems. We introduced various constrained coding schemes that address the sneak-path problem in these systems by removing data patterns that result in sneak paths. In particular, we designed RES-LOCO coding schemes defined over GF($4$) and GF($8$) where data is coded horizontally as well as an RLL coding scheme where data is coded vertically on the memristor crossbar array. We determined recursive formulae for the cardinalities of RES-LOCO codes. Moreover, we devised their encoding-decoding rules, where the codeword and its lexicographic index are bijectively related, which guarantee low complexity. We proposed bridging mechanisms that are rate-wise efficient. We presented experimental results that demonstrate the effectiveness of our coding schemes in mitigating the sneak-path problem. While RES-LOCO codes can remove the most detrimental sneak paths entirely when sequential reading is adopted, they still can be used with a variety of other circuit solutions and reading approaches. Future work includes incorporating memristor models and wire resistances in order to test our coding solutions under practical operating conditions. Another future direction is developing effective error-correction coding schemes for memristor crossbar arrays and combining them with our constrained coding solutions.

\section*{Acknowledgment}
The authors would like to thank Arash Motazedian for the helpful discussions on the research topic.



\begin{thebibliography}{13}


\bibitem{von_neumann}
J. Backus. ``Can programming be liberated from the von Neumann style?: A functional style and its algebra of programs,'' \textit{ACM Commun}, vol. 21, no. 8, pp. 613--641, Aug. 1978.

\bibitem{Chua1971Memristor}
L. Chua, ``Memristor-the missing circuit element,'' \textit{IEEE Trans. Circuit Theory}, vol. 18, no. 5, pp. 507--519, Sep. 1971.

\bibitem{Strukov2008Missing}
D. B. Strukov, G. S. Snider, D. R. Stewart, and R. S. Williams, ``The missing memristor found,'' \textit{Nature}, vol. 453, no. 7191, pp. 80--83, May 2008.

\bibitem{Chua2019Resistance}
L. Chua, ``Resistance switching memories are memristors,'' in \textit{Handbook of Memristor Networks}, L. Chua, G. Ch. Sirakoulis, and A. Adamatzky, Eds. Cham, Switzerland: Springer, 2019, pp. 197--230.

\bibitem{ref1}
A. Sebastian, M. Le Gallo, R. Khaddam-Aljameh, and E. Eleftheriou, ``Memory devices and applications for in-memory computing,'' \textit{Nat. Nanotechnol.}, vol. 15, no. 7, pp. 529--544, Jul. 2020.

\bibitem{Zidan2013Memristor}
M. A. Zidan, H. A. H. Fahmy, M. M. Hussain, and K. N. Salama, ``Memristor-based memory: The sneak paths problem and solutions,'' \textit{Microelectron. J.}, vol. 44, no. 2, pp. 176--183, Feb. 2013.

\bibitem{Zahoor2020Resistive}
F. Zahoor, T. Z. Azni Zulkifli, and F. A. Khanday, ``Resistive random access memory (RRAM): An overview of materials, switching mechanism, performance, multilevel cell (MLC) storage, modeling, and applications,'' \textit{Nanoscale Res. Lett.}, vol. 15, no. 1, p. 90, Apr. 2020.

\bibitem{Liang2013Effect}
J. Liang, S. Yeh, S. S. Wong, and H.-S. P. Wong, ``Effect of wordline/bitline scaling on the performance, energy consumption, and reliability of cross-point memory array,'' \textit{ACM J. Emerg. Technol. Comput. Syst.}, vol. 9, no. 1, pp. 9:1--9:14, Feb. 2013.

\bibitem{Lee2025Recent}
Y. Lee, B. Jeon, Y. Cho, J. Kim, W. Shim, and S. Kim, ``Recent progress in memristor array structures and solutions for sneak path current reduction,'' \textit{Adv. Mater. Technol.}, vol. 10, no. 4, p. 2400585, 2025.

\bibitem{ref3}
D. Ielmini and H.-S. P. Wong, ``In-memory computing with resistive switching devices,'' \textit{Nat. Electron.}, vol. 1, no. 6, pp. 333--343, Jun. 2018.

\bibitem{ref2}
A. Mehonic, A. Sebastian, B. Rajendran, O. Simeone, E. Vasilaki, and A. J. Kenyon, ``Memristors---from in-memory computing, deep learning acceleration, and spiking neural networks to the future of neuromorphic and bio-inspired computing,'' \textit{Adv. Intell. Syst.}, vol. 2, no. 11, p. 2000085, Aug. 2020.

\bibitem{Chen2020ReRAM}
Y. Chen, ``ReRAM: History, status, and future,'' \textit{IEEE Trans. Electron Devices}, vol. 67, no. 4, pp. 1420--1433, Apr. 2020.



\bibitem{Cassuto2014Channel}
Y. Cassuto, S. Kvatinsky, and E. Yaakobi, ``On the channel induced by sneak-path errors in memristor arrays,'' in \textit{Proc. Int. Conf. Signal Process. Commun. (SPCOM)}, Jul. 2014, pp. 1--6.

\bibitem{Cassuto2016Information}
Y. Cassuto, S. Kvatinsky, and E. Yaakobi, ``Information-theoretic sneak-path mitigation in memristor crossbar arrays,'' \textit{IEEE Trans. Inf. Theory}, vol. 62, no. 9, pp. 4801--4813, Sep. 2016.

\bibitem{Dupraz2023Turning}
E. Dupraz, F. Leduc-Primeau, K. Cai, and L. Dolecek, ``Turning to information theory to bring in-memory computing into practice,'' \textit{IEEE BITS Inf. Theory Mag.}, vol. 3, no. 3, pp. 64--77, Sep. 2023.

\bibitem{nguyen2023locally}
T. T. Nguyen, P. Li, K. Cai, and K. A. S. Immink, ``Locally mitigating sneak-path interference in resistive memory arrays,'' in \textit{Proc. IEEE Int. Symp. Inf. Theory (ISIT)}, Jun. 2023, pp. 1130--1135.

\bibitem{BenHur2019Detection}
Y. Ben-Hur and Y. Cassuto, ``Detection and coding schemes for sneak-path interference in resistive memory arrays,'' \textit{IEEE Trans. Commun.}, vol. 67, no. 6, pp. 3821--3833, Jun. 2019.

\bibitem{Pang2025Across}
Q. Pang and Z. Ma, ``Across-array LDPC codes design for resistive random-access memories,'' \textit{IEEE Trans. Comput.-Aided Des. Integr. Circuits Syst.}, 2025.

\bibitem{ref6}
B. Dai, K. Cai, Z. Mei, and X. Zhong, ``Polar code construction for resistive memories with sneak-path interference,'' \textit{IEEE Commun. Lett.}, vol. 28, no. 8, pp. 1765--1769, Aug. 2024.

\bibitem{Song2024Performance}
G. Song, M. Gao, Y. Li, B. Dai, and K. Cai, ``Performance analysis and code design for resistive random-access memory using channel decomposition approach,'' \textit{IEEE Trans. Inf. Theory}, vol. 72, no. 1, pp.  358--373, Jan. 2026.



\bibitem{Naous2014Memristor}
R. Naous, M. A. Zidan, A. Sultan-Salem, and K. N. Salama, ``Memristor based crossbar memory array sneak path estimation,'' in \textit{Proc. Int. Workshop Cellular Nanoscale Netw. Their Appl. (CNNA)}, Jul. 2014, pp. 1--2.

\bibitem{Chen2019Pilot}
Z. Chen, C. Schoeny, and L. Dolecek, ``Pilot assisted adaptive thresholding for sneak-path mitigation in resistive memories with failed selection devices,'' \textit{IEEE Trans. Commun.}, vol. 68, no. 1, pp. 66--81, Jan. 2020.

\bibitem{Pallathuvalappil2025Rate}
S. Pallathuvalappil and A. James, ``Rate coding with 3D memristor crossbar,'' \textit{IEEE Trans. Circuits Syst. Artif. Intell.}, vol. 2, no. 1, pp. 25--36, Mar. 2025.

\bibitem{Kim2022Sneak}
M. Kim and J. Ha, ``Sneak path aware bit-flipping algorithm for ReRAM crossbar array,'' in \textit{Proc. Int. Conf. Inf. Commun. Technol. Converg. (ICTC)}, Oct. 2022, pp. 451--454.

\bibitem{Kong2025Sneak}
L. Kong, Y. Qi, H. Liu, and C. Meng, ``Sneak path-aware reliability-based iterative majority-logic decoding algorithms for LDPC codes in ReRAM systems,'' \textit{IEEE Commun. Lett.}, vol. 29, no. 9, pp. 2018--2022, Sep. 2025.

\bibitem{Sun2022Belief}
C. Sun, K. Cai, G. Song, T. Q. S. Quek, and Z. Fei, ``Belief propagation based joint detection and decoding for resistive random access memories,'' \textit{IEEE Trans. Commun.}, vol. 70, no. 4, pp. 2227--2239, Apr. 2022.



\bibitem{shan_const}
C. E. Shannon, ``A mathematical theory of communication,'' Bell Sys. Tech. J., vol. 27, Oct. 1948.

\bibitem{tang_bahl}
D. T. Tang and R. L. Bahl, ``Block codes for a class of constrained noiseless channels,'' \emph{Inf. and Control}, vol. 17, no. 5, pp. 436--461, 1970.

\bibitem{datta_2}
{S. Datta and S. W. McLaughlin, ``Optimal block codes for M-ary runlength-constrained channels,'' \emph{IEEE Trans. Inf. Theory}, vol. 47, no. 5, pp. 2069--2078, Jul. 2001.}

\bibitem{Hareedy2019Asymmetric}
A. Hareedy and R. Calderbank, ``Asymmetric LOCO codes: Constrained codes for flash memories,'' in \textit{Proc. Annu. Allerton Conf. Commun., Control, Comput.}, Sep. 2019, pp. 124--131.

\bibitem{ahh_qaloco}
A. Hareedy, B. Dabak, and R. Calderbank, ``Managing device lifecycle: Reconfigurable constrained codes for M/T/Q/P-LC Flash memories,'' \emph{IEEE Trans. Inf. Theory}, vol. 67, no. 1, pp. 282--295, Jun. 2021.

\bibitem{wang_etal}
Y. Wang, M. Noor-A-Rahim, E. Gunawan, Y. L. Guan, and C. L. Poh, ``Construction of bio-constrained code for DNA data storage,'' \textit{IEEE Commun. Lett.}, vol. 23, no. 6, pp. 963--966, Jun. 2019.

\bibitem{Hareedy2019LOCO}
A. Hareedy and R. Calderbank, ``LOCO codes: Lexicographically-ordered constrained codes,'' \textit{IEEE Trans. Inf. Theory}, vol. 66, no. 6, pp. 3572--3589, Jun. 2020.

\bibitem{cover_lex}
T. Cover, ``Enumerative source encoding,'' \emph{IEEE Trans. Inf. Theory}, vol. 19, no. 1, pp. 73--77, Jan. 1973.

\bibitem{Hareedy2022LOCO}
A. Hareedy, B. Dabak, and R. Calderbank, ``The secret arithmetic of patterns: A general method for designing constrained codes based on lexicographic indexing,'' \textit{IEEE Trans. Inf. Theory}, vol. 68, no. 9, pp. 5747--5778, Sep. 2022.

\bibitem{Ozbayrak2024TDLOCO}
I. Guzel, D. \"{O}zbayrak, R. Calderbank, and A. Hareedy, ``Eliminating media noise while preserving storage capacity: Reconfigurable constrained codes for two-dimensional magnetic recording,''  \textit{IEEE Trans. Inf. Theory}, vol. 70, no. 7, pp. 4905--4927, Jul. 2024.

\bibitem{Irimagzi2024Protecting}
C. \.{I}rima\u{g}z\i{}, Y. Uslan, and A. Hareedy, ``Protecting the future of information: LOCO coding with error detection for DNA data storage,'' \textit{IEEE Trans. Mol., Biol., Multi-Scale Commun.}, vol. 10, no. 2, pp. 317--333, Jun. 2024.

\bibitem{reins_EC_DLOCO}
C. \.{I}rima\u{g}z\i{} and A. Hareedy, ``LOCO codes can correct as well: Error-correction constrained coding for DNA data storage,'' \textit{IEEE Trans. Commun.}, vol. 74, pp. 2235-2250, Jan. 2026.

\bibitem{marcus_book}
B. H. Marcus, R. M. Roth, and P. H. Siegel, \textit{An Introduction to Coding for Constrained Systems.} Lecture notes, 2001.

\bibitem{ahh_simpleloco}
D. \"{O}zbayrak, D. Uyar, and A. Hareedy, ``Low-Complexity Constrained Coding Schemes for Two-Dimensional Magnetic Recording,'' in \emph{Proc. IEEE Int. Symp. Inf. Theory (ISIT)}, Athens, GR, Jul. 2024, pp. 1--6.

\end{thebibliography}
\end{document}